\documentclass[journal]{IEEEtran}

\usepackage{xcolor,soul,framed} 

\colorlet{shadecolor}{yellow}
\usepackage[pdftex]{graphicx}
\graphicspath{{../pdf/}{../jpeg/}}
\DeclareGraphicsExtensions{.pdf,.jpeg,.png}

\usepackage[cmex10]{amsmath}
\usepackage{array}
\usepackage{eqparbox}
\usepackage{url}

\usepackage{diagbox}
\usepackage{booktabs}
\usepackage{multirow}
\usepackage{tabularx}
\usepackage{array}
\usepackage{makecell}
\usepackage{pifont}   
\usepackage{threeparttable}
\usepackage{xcolor}   
\usepackage{colortbl}
\newcommand{\graycmidrule}[1]{%
  \arrayrulecolor{gray!30}\cmidrule(lr){#1}\arrayrulecolor{black}}
\newcolumntype{L}[1]{>{\raggedright\arraybackslash}p{#1}}
\newcommand{\cmark}{\ding{51}} 
\newcommand{\xmark}{\ding{55}} 

\begin{document}
\title{Parameter Estimation of Power Electronic Converters with Differentiable Physics Simulation} \author{Pingjunjin~Tan,~\IEEEmembership{Student Member,~IEEE,} Chunlin~Lv,~\IEEEmembership{Member,~IEEE,} Jinjun~Liu,~\IEEEmembership{Fellow,~IEEE,} and~Yang~Li,~\IEEEmembership{Student Member,~IEEE}

}

\markboth{ 
}{ }

\maketitle

\begin{abstract}
This article proposes a differentiable physics simulation (DP simulation)-based parameter estimation method for the condition monitoring of power electronic converters. In the proposed method, the time-domain simulation of converter dynamics is embedded into a differentiable computational graph, directly linking device parameters to observed voltage and current trajectories. By formulating differentiable time-stepping operators, the nonlinear dynamics of the converter across different circuit topologies are simulated in a unified, differentiable manner. A dc–-dc buck converter is used as a representative case study. Using sparse transient samples from existing sensing channels, the method enables noninvasive parameter estimation without additional sensing hardware. Comprehensive simulation studies are conducted to evaluate the impacts of time-stepping schemes, regularization constraints, and various uncertainty sources on estimation accuracy and robustness. Subsequently, 30 distinct hardware configurations are experimentally tested for validation. The results show that the proposed method can effectively track the relative variations of health-related parameters across the critical components. This DP simulation framework provides a novel perspective for physics-informed machine learning in power electronic applications.
\end{abstract}

\begin{IEEEkeywords}
buck converter, condition monitoring, differentiable physics simulation, physics-informed machine learning, predictive maintenance
\end{IEEEkeywords}

%
\IEEEpeerreviewmaketitle


\section{Introduction}

\IEEEPARstart{S}{witched-mode} power supplies (SMPS) have been widely deployed across diverse applications over the past three decades. As service time increases, components progressively degrade, resulting in reduced electrothermal stress tolerance and higher maintenance requirements \cite{tan2025robust}. Consequently, reliability has emerged as a considerable practical challenge in power electronics, particularly in safety-critical applications such as electrified transportation, renewable energy generation, and data centers. Failure of key components in SMPSs can result in severe safety hazards and substantial economic losses. According to industrial surveys reported in \cite{Survey2011} and \cite{Survey2018}, condition monitoring is expected to become a crucial tool in reliability engineering by tracking the degradation of fragile components and enabling timely intervention before catastrophic failures occur. 

Component aging often stems from micro-scale deterioration of materials and structures that is difficult to observe directly \cite{ConMon2010}. Fortunately, micro-scale degradation manifests as measurable macroscopic changes in electrical, thermal, and structural behavior. In practice, most fielded power electronic converters are only instrumented with electrical sensors. Therefore, an implementation-friendly approach for condition monitoring is to track the variations of intrinsic electrical parameters of critical components as health indicators, such as the on-state resistance of MOSFET and capacitance value of capacitor. 

\begin{figure}
  \begin{center}
  \includegraphics[width=3.5in]{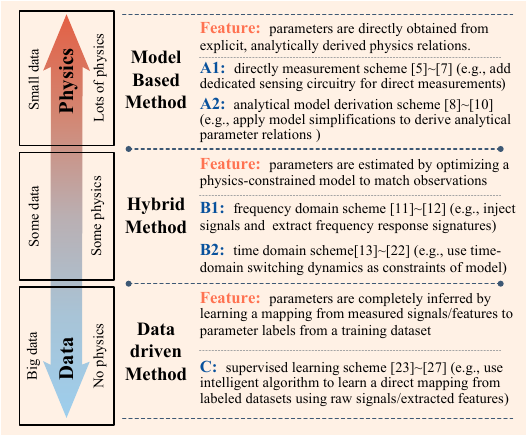}\\
  \caption{Classification of typical parameter estimation methods for power electronic converter.}\label{fig:classification}
  \end{center}
\end{figure}

In recent years, significant progress has been made in online estimation of health indicators for critical components in SMPSs. From the perspective of the degree of physics model constraint and data-driven integration, they can be categorized into three types: model-based \cite{Rogowski2022Zxm}, \cite{Novel2024Asoodar}, \cite{Accurate2023Asoodar}, \cite{Online2021Zzy}, \cite{ESR2020Lwg}, \cite{IGBTCap2017Spj}, hybrid \cite{In2016Serkan}, \cite{Low2016}, \cite{Nonlinear2021Rojas}, \cite{Parameter2018Riba}, \cite{Digital2021Pyz}, \cite{Electrothermal2023}, \cite{OneCycle2024Cgp}, \cite{Digital2021chen}, \cite{Condition2017}, \cite{Parameter2022Zs}, \cite{Physics2025Osy}, \cite{Extended2025Xyx}, and data-driven \cite{Full2024Cgp}, \cite{Non-Invasive2019}, \cite{2023Parameter}, \cite{Deep2022Park}, \cite{Artificial2017Soliman}. Fig.~\ref{fig:classification} outlines the key features and subcategories of these three types of methods. The typical implementation schemes for parameter estimation of power electronic converters are further summarized in Table~\ref{tab:cm_conveter}.

\begin{table*}[t]
\caption{Typical Implementation Schemes for Parameter Estimation of Power Converters}
\label{tab:cm_conveter}
\centering
\scriptsize
\setlength{\tabcolsep}{3pt}
\renewcommand{\arraystretch}{1}

\resizebox{\textwidth}{!}{
\begin{threeparttable}
\begin{tabular}{c c l l l l l l m{5.5cm}}
\toprule
\makecell[l]{\textbf{Type}} &
\makecell[l]{\textbf{Ref}} &
\makecell[l]{\textbf{Topology}} &
\makecell[l]{\textbf{Target}\\\textbf{Component}}&
\makecell[l]{\textbf{$f_{sw}$ (N)}} &
\makecell[l]{\textbf{Used}\\\textbf{Signal}} &
\makecell[l]{\textbf{Additional Sampling}\\\textbf{Circuits~/~Devices}} &
\makecell[l]{\textbf{Estimation}~\textbf{Error}} &
\makecell[l]{\textbf{Advantages and limitations}}\\
\midrule

\multirow{6}{*}{A1} & {\cite{Rogowski2022Zxm}} & Inverter & Cap. & 12.5 kHz (N = 8) & $v_{dc},\ i_{cap}$ & Rogowski coil sensor & $R_C$: $<2.8\%$ & \makecell[l]{\cmark\ Realize individual capacitor monitoring\\
\xmark\ Specially designed current sensor}\\
\graycmidrule{2-9}

& {\cite{Novel2024Asoodar}}  & Inverter & Mos. & 10.11 kHz (N/A) & $v_{ds},\ i_s,\ T_{sink}$ & \makecell[l]{on-state voltage \\measurement circuit} & $R_{dson}$: $<1\%$ & \multirow{4}{*}{\makecell[l]{\cmark\ High accuracy, temperature effects considered\\
\xmark\ Extra dedicated sampling circuit}}\\
\graycmidrule{2-8}

& {\cite{Accurate2023Asoodar}} & MMC & Mos.  & N/A & $v_{ds},\ i_s,\ T_{sink}$      &
\makecell[l]{on-state voltage \\measurement circuit} & $R_{dson}$: $<1.5\%$ &  \\
\midrule

\multirow{7}{*}{A2}& {\cite{Online2021Zzy}} & Boost & Cap. & 200 kHz ($N=1$) & $v_{o},i_{o},d$& Load current sensor &
C: $<3\%$ &
\makecell[l]{\cmark\ Low sampling frequency\\
\xmark\ Additional sensors and circuits, long transient \\ period data needed} \\
\graycmidrule{2-9}

& {\cite{ESR2020Lwg}} & Boost PFC & Cap. & 60 kHz ($N=53.3$) & $v_o,i_L,$ &
\makecell[l]{Voltage ripples \\extraction circuits} &
\makecell[l]{$R_C$: $<10\%$} &\makecell[l]{
\cmark\ No additional current sensor\\
\xmark\ Extra hardware circuits, high frequency sampling} \\
\graycmidrule{2-9}

& {\cite{IGBTCap2017Spj}} & Inverter & IGBT, Cap. & 4 kHz ($N/A$) & $v_{dc},i_{sc}$ &
N/A &
\makecell[l]{$R_C$: $<6\%$\\ IGBT:(unspecific)} &\makecell[l]{
\cmark\ No additional hardware circuits\\
\xmark\ Require special operating condition, quasi-online} \\
\midrule

\multirow{3}{*}{B1}& {\cite{In2016Serkan}} & Boost & Mos. & 100 kHz (N/A) & $v_o,i_L$ & Not required &
$R_{dson}$: $<5\%$ &\multirow{3}{*}{\makecell[l]{
\cmark\  No extra measurement circuit; low-sampling and\\ multiparameter estimation~\cite{Low2016}\\
\xmark\ Require signal~/~PRBS injection}}\\
\graycmidrule{2-8}

& {\cite{Low2016}} & Buck  & Ind., Cap. & 20 kHz (N$\approx$0.14)  & $i_L,v_o,d$   &
Not required &
\makecell[l]{L: $<5.2\%$\\ C: $<7.5\%$, $R_C$: $<0.4\%$} &
\\
\midrule


\multirow{32}{*}{B2} &
{\cite{Nonlinear2021Rojas}} & \makecell[l]{Buck, Boost\\Buck-boost}& \makecell[l]{Mos.,\\Ind., Cap.} & \makecell[l]{200~/~500 kHz \\($N>10000$)} & $i_{in},v_{in},v_o$ &
\makecell[l]{Oscilloscope (HF sampling)} & \makecell[l]{C: $<17.5\%$,$R_C$: $<31\%$\\L: $<7.3\%$,$R_L$: $<21.9\%$} &
\multirow{4}{*}{\makecell[l]{
\cmark\ Multiparameter estimation, including components \\and control-loop parameters\\
\xmark\ High sampling frequency (up to GHz), commercial \\simulator or toolbox needed}}\\
\graycmidrule{2-8}

& {\cite{Parameter2018Riba}} & Buck, Boost & \makecell[l]{Mos.,\\Ind., Cap.} & \makecell[l]{200~/~480 kHz \\($N>5000$)} & $i_{in},v_{in},v_o,i_o$ &
\makecell[l]{Oscilloscope (HF sampling)}& \makecell[l]{C: $<15.1\%$,$R_C$: $<3.4\%$\\L: $<7.0\%$,$R_L$: $<21.9\%$} &
 \\
\graycmidrule{2-9}

& {\cite{Digital2021Pyz}} & Buck & \makecell[l]{Mos.,\\Ind., Cap.} & 20 kHz (N=2.5) & $v_{in},\ v_{o},\ i_L,\ s$ &
Not required & 
\makecell[l]{C: $<5.7\%$} & 
\multirow{10}{*}{
\makecell[l]{\cmark\ Multiparameter estimation, noninvasive, moderate \\computational cost; low sampling frequency~\cite{Digital2021Pyz}\cite{OneCycle2024Cgp}\\ \cite{Digital2021chen}, electrothermal effects considered~\cite{Electrothermal2023}\\
\\
\xmark\ Well-defined search ranges and hyperparameters \\needed; require transient data \cite{Digital2021Pyz},  simulation-only \\validation \cite{Digital2021chen}, commercial simulator needed~\cite{OneCycle2024Cgp}\cite{Digital2021chen}}}
 \\
\graycmidrule{2-8}

& {\cite{Electrothermal2023}} & Buck & \makecell[l]{Mos., Cap.} & 100 kHz (N=100) & $v_{in},\ v_{o},\ i_L,\ s$ &
\makecell[l]{Oscilloscope (HF sampling)} &
\makecell[l]{C: $<6\%$, $R_C$: $<0.8\%$\\$R_{dson}$:$<0.6\%$} &
\\
\graycmidrule{2-8}

& {\cite{OneCycle2024Cgp}} & Buck & \makecell[l]{Mos.,\\Ind., Cap.} & 20 kHz (N=$4\sim8$)      & $v_{in},\ v_{o},\ i_L, s$ &
Not required &
\makecell[l]{C: $<1.3\%$, $R_C$: $<5.4\%$\\L: $<2.7\%$, $R_L$: $<1.5\%$} &
 \\
\graycmidrule{2-8}

& {\cite{Digital2021chen}} & Boost & \makecell[l]{Mos., Dio.\\Ind., Cap.} & 20 kHz ($N=5$) & $v_{in},\ v_{o},\ i_L$ &
Not required &
\makecell[l]{C: $<0.48\%$, $R_C$: $<5\%$\\L: $<0.7\%$, $R_L$: $<20\%$\\$R_{on}$: $<20\%$, $R_{dson}$: $<44.8\%$} &
 \\
\graycmidrule{2-9}

& {\cite{Condition2017}} & Buck & \makecell[l]{Ind., Cap.} & 300 kHz ($N/A$)     & $v_{in},\ v_{o},\ i_L, s$ &
NI CompactRIO &
\makecell[l]{C: $<5\%$,\\L: $<5\%$} &
\makecell[l]{
\cmark\ Multiparameter estimation, fault detection available\\
\xmark\ Hyperparameter sensitive} \\
\graycmidrule{2-9}

& {\cite{Parameter2022Zs}} & Buck & \makecell[l]{Mos., Dio.\\Ind., Cap.} & 20 kHz ($N=2$)     & $ v_{o},\ i_L, s$ &
Not required &
\makecell[l]{C: $<1.6\%$, $R_C$: $<11.5\%$,\\ $R_{dson}+R_L$: $<18.5\%$}& \multirow{4}{*}{\makecell[l]{
\cmark\ No labeled pre-training, data-light\\
\xmark\ high computational cost, hard to converge, require\\ transient data; limited peak-sampling accuracy \cite{Parameter2022Zs},\\simulation-only validation \cite{Physics2025Osy} } }
\\
\graycmidrule{2-8}

& {\cite{Physics2025Osy}} & Buck & \makecell[l]{Mos., Dio.\\Ind., Cap.} & 20 kHz ($N=2$)     & $ v_{in}, v_{o}, i_L, s$ &
Not required &
\makecell[l]{C: $<5\%$, $R_C$: $<11\%$\\L: $<7\%$}&
\\
\graycmidrule{2-9}

& {\cite{Extended2025Xyx}} & Buck & \makecell[l]{Mos., Dio.\\Ind., Cap.} & 20 kHz ($N=2$)     & $  v_{o}, i_L, s$ &
Not required &
\makecell[l]{C: $<1.1\%$, $R_C$: $<0.8\%$\\L: $<4.4\%$, $V_{F}$: $<19.1\%$}&
\makecell[l]{
\cmark\ Multiparameter estimation, applicable to DCM\\
\xmark\ higher computational cost} \\
\midrule


\multirow{10}{*}{C}& {\cite{Full2024Cgp}} & Buck & \makecell[l]{Mos., Dio.\\Ind., Cap.} & 20 kHz ($N=10$)     & $  v_{o}, i_L$ &
N/A &
\makecell[l]{C: $<3.0\%$, $R_C$: $<1.7\%$\\L: $<1.6\%$, $R_L$: $<4.9\%$,\\$\Delta R_{dson}$: $<2.5\%$}&\makecell[l]{
\cmark\ Very high accuracy, multiparameter estimation,\\reduced training data demand \\
\xmark\ Need pre-training, specific topology}
 \\
\graycmidrule{2-9}

& {\cite{Non-Invasive2019}} & STATCOM & \makecell[l]{Ind., Cap.} & N/A    & $  v_{pcc},i_s,i_t$ &
Power Quality Analyzer &
\makecell[l]{$\mathrm{C_{dc}}$: $<0.12\%$\\$L_s$: $<4.2\%$, $R_s$: $<0.062\%$,\\$L_f$: $<7.5\%$, $R_f$: $<1.56\%$}&
\multirow{7}{*}{\makecell[l]{
\cmark\ Very high accuracy, multiparameter estimation,\\ fast NN-based inference, only simulation training \\data needed \cite{Non-Invasive2019}\\
\\
\xmark\ Need pre-training, features preprocessing needed, \\specific topology, high sampling frequency}}
 \\
\graycmidrule{2-8}

& {\cite{2023Parameter}} & Buck & \makecell[l]{Ind., Cap.} & 40 kHz ($N=25$)     & $v_{o}$ &
N/A &
\makecell[l]{C: $<2.1\%$, $R_C$: $<2.7\%$\\L: $<2.5\%$, $R_L$: $<2.9\%$}&
\\
\graycmidrule{2-8}

& {\cite{Deep2022Park}} & Inverter & \makecell[l]{ Cap.} & 5 kHz ($N=50$)     & $  v_{C},i_C$ &
Oscilloscope (HF sampling) &
\makecell[l]{C: $<0.11\%$, $R_C$: $<0.11\%$}&
\\
\graycmidrule{2-8}

& {\cite{Artificial2017Soliman}} & Inverter & \makecell[l]{ Cap.} & N/A     & $  i_{o},\Delta v_{dc}$ &
Oscilloscope (HF sampling) &
\makecell[l]{C: $<0.5\%$}&
\\
\midrule

\multicolumn{2}{c}{\makecell[c]{Proposed\\method}}
& Buck 
& \makecell[l]{Mos., Dio.\\Ind., Cap.} 
& 20 kHz ($N=2$) 
& $v_o,\ i_L,\ s$ 
& Not required 
& \makecell[l]{Exp. variation deviation:\\
$C<3.2\%,\ R_C<18.8\%$\\
$L<4.5\%$,\ $V_F:$ poor accuracy\\
$R_{\rm dson}+R_L<12.2\%$
}
& \makecell[l]{
\cmark\ Noninvasive, low sampling burden, fast convergence,\\
\quad multiparameter estimation, physically interpretable\\
\xmark\ Require transient data and known switching instants}
\\

\bottomrule
\end{tabular}
\begin{tablenotes}[flushleft]
\footnotesize
\item[] \textit{Note:} Errors are reported according to the metrics used in the original papers and are not directly comparable due to different test conditions, parameter definitions, and validation platforms. (Mos. = MOSFET, Dio. = Diode, Ind. = inductor, Cap. = capacitor; N/A indicates not available or not applicable in the original paper.)
\end{tablenotes}
\end{threeparttable} 
}
\end{table*}

Ideally, the most effective strategy for parameter estimation of SMPS involves establishing explicit analytical mappings between component parameters and observed signals. However, constructing such expressions is non-trivial due to the high nonlinearity inherent in SMPS operation. As detailed in Table~\ref{tab:cm_conveter}, direct measurement schemes (i.e., category A1) address this challenge by employing auxiliary sensing circuits to capture terminal information, such as customized Rogowski coils for capacitor currents~\cite{Rogowski2022Zxm} or isolated detection circuits for the drain-source voltage of power semiconductors~\cite{Novel2024Asoodar},~\cite{Accurate2023Asoodar}. While intuitively straightforward and universally applicable, these extra sensing circuits introduce potential failure points, thereby compromising system reliability and complicating integration into mature converter designs.

Alternatively, another category of model-based strategies (i.e., A2) leverages simplified models~\cite{Online2021Zzy},~\cite{ESR2020Lwg} or specific operating conditions~\cite{IGBTCap2017Spj} to deduce parameters without additional sensors. Although this reduces hardware overhead, such methods often necessitate the assumption that other circuit parameters remain constant and known. However, in practical scenarios, degradation could occur concomitantly across multiple components~\cite{PERState2021}. Therefore, given the 'weakest-link' nature of reliability, it is critical to prioritize whole-converter monitoring over isolated component estimation.

In contrast to model-based methods, data-driven methods (i.e., category C) estimate parameters without any prior knowledge of circuit topology, control strategies, or operating modes. Extensive research has focused on dataset construction (e.g., synthetic data and experimental measurements) \cite{Full2024Cgp}, \cite{Non-Invasive2019} and feature extraction (e.g., time-domain and frequency-domain features)~\cite{2023Parameter},~\cite{Deep2022Park},~\cite{Artificial2017Soliman}. Once trained, these models enable single-pass inference, facilitating low computational costs and execution time. However, purely data-driven methods face two inherent limitations in power electronics. First, unlike image recognition or natural language processing, power electronics remains a data-scarce field~\cite{Datadriven2021Zs}. Second, despite their high accuracy shown in Table~\ref{tab:cm_conveter}, these methods struggle with data distribution shifts. Mismatches between training and testing data can lead to catastrophic prediction errors. Such unreliable predictions, owing to extrapolation or observational biases, may be physically inconsistent~\cite{PhyML2021LuLu} and therefore unacceptable for practical engineering.

To mitigate the aforementioned limitations, hybrid methods have emerged as a promising compromise. Depending on the type of embedded physical constraints, these hybrid methods are categorized into frequency-domain response and time-domain dynamics schemes. In frequency-domain schemes (i.e., Category B1)~\cite{In2016Serkan},~\cite{Low2016}, perturbation signals like Pseudo-Random Binary Sequences (PRBS) are injected into the control loop to extract frequency characteristics. However, this method hinges on an indirect inference chain, where errors in transfer function identification propagate to the final parameters. In addition, the required signal injection may compromise system stability.

Conversely, time-domain strategies (i.e., Category B2) directly leverage the converter's inherent switching dynamics as physical constraints. These methods typically construct a parallel physical model to track the actual hardware, encompassing implementations such as circuit simulators~\cite{Nonlinear2021Rojas},~\cite{Parameter2018Riba}, numerical propagation~\cite{Digital2021Pyz},~\cite{Electrothermal2023},~\cite{OneCycle2024Cgp}, \cite{Digital2021chen}, adaptive observers~\cite{Condition2017}, or physics-informed neural networks (PINNs) \cite{Parameter2022Zs}, \cite{Physics2025Osy}, \cite{Extended2025Xyx}. Minimizing the deviation between the model-predicted trajectories and measured sensor data subsequently enables accurate parameter identification. In \cite{Nonlinear2021Rojas} and \cite{Parameter2018Riba}, commercial simulation platforms (e.g., MATLAB/Simulink and PSIM) serve as the prerequisite for model construction and data generation. This reliance limits deployment flexibility, compounded by the prohibitive time overhead from mandatory model recompilation. 
Alternatively, adaptive observers \cite{Condition2017} utilize Lyapunov stability theory to derive update laws, allowing for real-time parameter convergence based on state estimation errors. Consequently, parasitic elements are typically ignored to avoid the excessive complexity associated with high-dimensional gain matrix design.

To circumvent these modeling complexities, a more direct strategy involves explicitly reconstructing the discrete-time evolution dynamics of the converter via classical numerical analysis. In~\cite{Digital2021Pyz},~\cite{Electrothermal2023},~\cite{OneCycle2024Cgp}, and~\cite{Digital2021chen}, explicit Runge--Kutta methods are utilized to formulate the analytical relationship between key electrical parameters and voltage/current trajectories. Subsequently, derivative-free optimization algorithms, such as Particle Swarm Optimization (PSO) and Bayesian Optimization (BO), are employed to iteratively search for optimal parameters by minimizing the model prediction error. While this paradigm boasts excellent versatility across arbitrary topologies, these derivative-free search mechanisms typically necessitate extensive function evaluations. Consequently, they incur a prohibitive computational burden in high-dimensional parameter identification tasks and exhibit extreme sensitivity to hyperparameter settings \cite{talbi2009metaheuristics}.

Similar to numerical propagation scheme \cite{Digital2021Pyz}, \cite{Electrothermal2023}, \cite{OneCycle2024Cgp}, \cite{Digital2021chen}, PINNs formulate physical constraints via explicit \cite{Physics2025Osy} or implicit \cite{Parameter2022Zs}, \cite{Extended2025Xyx} Runge--Kutta discretization. These constraints are then explicitly embedded into the training loss. Since the loss function is differentiable with respect to the neural network weights and the embedded physical parameters, the training process can be optimized via gradient-based algorithms. Compared to heuristic searches, gradient-based methods leverage directional derivative information to achieve significantly faster convergence and higher computational efficiency. However, the Runge--Kutta constraints used in existing PINN-based formulations are typically constructed for a single continuous dynamical model~\cite{PINN2019Raissi}. In switching converters, such constraints are therefore valid only within topology-invariant intervals and cannot naturally relate samples across switching transitions. To maximize data efficiency, sampling is therefore forced to align with switching instants \cite{Parameter2022Zs}. Unfortunately, measurements at these transitions are inherently susceptible to high-frequency switching noise and ringing. Furthermore, the neural network architecture introduces a vast parameter space. Co-optimizing these weights alongside the target component parameters creates a highly non-convex landscape that severely compromises convergence stability \cite{Understanding2021Wang}.

In recent years, the concept of Physics-Informed Machine Learning (PIML) has garnered increasing attention, with deep learning techniques being a primary focal point. However, classical numerical solvers have evolved over decades, establishing a mature and rigorous theoretical foundation for system simulation. It is imprudent to disregard this accumulated domain knowledge by blindly adopting generic deep learning frameworks \cite{DBLP}. Acting as a bridge between scientific computing and modern deep learning, Differentiable Physics (DP) offers a transformative solution \cite{ramsundar2021differentiable}. By reformulating the discrete numerical integration into a differentiable computational graph, this paradigm enables end-to-end gradient backpropagation directly through the physical evolution dynamics. Intrinsic circuit parameters can be optimized via gradient descent without relying on neural network surrogates, ensuring both physical consistency and computational efficiency.

This article aims to deliver these benefits of DP to the power electronic field. As an exemplary application, it proposes a new method of parameter estimation of a dc–-dc buck converter based on DP simulation. The code and data accompanying this article are available in the Supplementary Material and also on a GitHub repository\footnote{Available online: https://github.com/TPjj}. The presented study has the following contributions and advantages.
\begin{enumerate}
    \item The proposed framework reformulates numerical integration as a differentiable computational graph, enabling end-to-end parameter optimization directly through the converter dynamics.
    \item Seamless integration across topological transitions permits flexible sampling away from noise-prone switching instants, ensuring high accuracy even with sparse data.
    \item The framework realizes the simultaneous estimation of all critical circuit parameters, featuring a topology-agnostic design readily extensible to other architectures like dc--ac inverters.
\end{enumerate}

The rest of this article is organized as follows. Section II presents the methodology, including the physical model of a dc–-dc buck converter, idea and framework of the DP simulation, and data preparation procedure. Section III and Section IV verify the proposed method in the simulation and experiment testing, respectively. Finally, Section V concludes this article.

\section{Methodology}

\subsection {Dynamic Modeling of the Buck Converter in CCM}

\begin{figure}
  \begin{center}
  \includegraphics[width=3.5in]{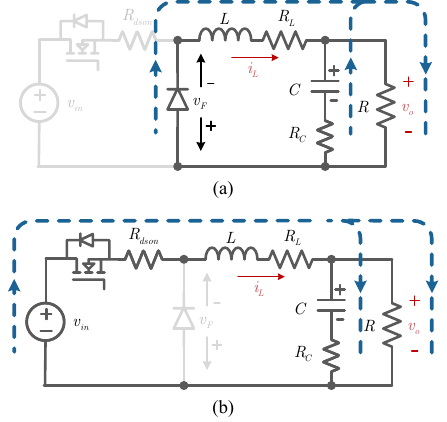}\\
  \caption{Equivalent circuits of the buck converter under two switching states. (a) MOSFET off-state operation; (b) MOSFET on-state operation.}\label{circuit}
  \end{center}
\end{figure}

The inherent nonlinearity in the dynamic behavior of power electronic converters primarily stems from the periodic alteration of circuit topologies dictated by semiconductor switches. The dc--dc buck converter serves as the most fundamental and ubiquitous topology in power electronics. Operating in Continuous Conduction Mode (CCM), the switching actions of the MOSFET divide the converter into two distinct linear circuit stages, depicted in Fig.~\ref{circuit} (a) and Fig.~\ref{circuit} (b), respectively.

Fundamentally, the Pulse Width Modulation (PWM) technique employed to regulate the converter governs these semiconductor switching behaviors. To mathematically unify the system dynamics across different topologies, a discrete switching function $S \in \{0,1\}$ is introduced to represent the on/off state of the active switch. In a buck converter, the inductor current $i_L$ and output voltage $v_o$ are typically sampled and fed back to the control loop. Consequently, considering all relevant parasitic parameters and control loop design, the comprehensive state-space formulation of the buck converter can be expressed as:

\begin{equation}\label{state_space_eq}
\begin{array}{c}
\displaystyle
\begin{bmatrix}
{d i_L}/{dt} \\[4pt]
{d v_o}/{dt}
\end{bmatrix}
=
\mathbf{M}_S
\begin{bmatrix}
i_L \\[3pt]
v_o
\end{bmatrix}
+
\mathbf{d}_S,\\[30pt]
\displaystyle
\mathbf{M}_S=
\begin{bmatrix}
-\dfrac{SR_{dson}+R_L}{L} & -\dfrac{1}{L} \\[8pt]
\dfrac{LR-CRR_C(SR_{dson}+R_L)}{LC(R+R_C)} &
-\dfrac{CRR_C+L}{LC(R+R_C)}
\end{bmatrix}, \\[30pt]
\displaystyle
\mathbf{d}_S=
\begin{bmatrix}
\dfrac{S}{L}v_{in}-\dfrac{1-S}{L}v_F \\[8pt]
\dfrac{RR_C}{R+R_C}
\left(
\dfrac{S}{L}v_{in}-\dfrac{1-S}{L}v_F
\right)
\end{bmatrix},
\end{array}
\end{equation}

\noindent
where the inductor current $i_L$ and output voltage $v_o$ are selected as state variables to directly align with the sparsely sampled observational data. $v_{in}$ denotes the input DC voltage, and $v_F$ represents the forward voltage drop of the freewheeling diode. $R_{dson}$ signifies the MOSFET on-state resistance. $L$ and $C$ denote the inductance and dc-link capacitance, alongside their respective equivalent series resistances $R_L$ and $R_C$. $R$ denotes the load resistance.

\subsection {Generalized Runge--Kutta Time-stepping Scheme}

The continuous-time state-space model presented in \eqref{state_space_eq} is a system of linear ordinary differential equations (ODEs). Given the initial conditions of the inductor current and output voltage, an analytical closed-form solution can theoretically be obtained. However, deriving this exact analytical expression necessitates complex algebraic operations, such as eigendecomposition and matrix inversion. Such analytical forms are highly rigid and intractable to be formulated in a manner suitable for converter parameter identification.

A more practical alternative is to employ a time-stepping strategy based on numerical analysis methods. This approach transforms the continuous dynamic behavior into a time-discrete representation within an acceptable accuracy margin. Formally, consider a dynamic system governed by a general ordinary differential equation:

\begin{equation}\label{general_ODEs}
\mathbf{u}'=f(t,\mathbf{u}),\quad t \geq t_0, \qquad \mathbf{u}(t_0)=\mathbf{u}_0 .
\end{equation}

As illustrated in \eqref{generic_step}, the unknown state vector at the $(n+1)$th time step $\mathbf{u}_{n+1}$ can be iteratively deduced from the known state vector $\mathbf{u}_n$ at the $n$th step. $h$ is the numerical integration step size and $\Phi$ represents a generic increment function determined by the specific numerical method.

\begin{equation}\label{generic_step}
\mathbf{u}_{n+1}=\mathbf{u}_n+h\,\Phi(\mathbf{u}_n,t_n,h,f) .
\end{equation}

There is a multitude of methods for the numerical solution of the ODE system. The Runge--Kutta (RK) family of methods stands out as one of the most widely adopted numerical algorithms. The idea behind RK methods is to replace an integral with a finite sum. They evaluate the system states at multiple intermediate stages within the step interval. A weighted algebraic sum of these approximations then yields the numerical solution. The scheme can be derived as:

\begin{equation}\label{eq_runge_kutta}
\begin{aligned}
\mathbf{\xi}_j &= \mathbf{u}_n + h \sum_{i=1}^{\nu} a_{j,i}\, f(t_n + c_i h, \mathbf{\xi}_i),\\
\mathbf{u}_{n+1} &= \mathbf{u}_n + h \sum_{j=1}^{\nu} b_j\, f(t_n + c_j h, \mathbf{\xi}_j),
\end{aligned}
\end{equation}

\noindent
where $\nu$ is the number of stages in the RK scheme. $\mathbf{\xi}_j$ is the intermediate state approximation at each respective stage. The matrix $A=(a_{i,j})$ is called the RK coefficient matrix, while $b_j$ and $c_j$ are the RK weights and RK nodes respectively \cite{iserles2009first}. All of these coefficients can be displayed in a Butcher tableau with the following form:

\begin{equation}\label{eq_butcher_tableau}
\left.
\begin{array}{c|c}
c & A\\
\hline
 & \raisebox{-2pt}{$b^T$}
\end{array}
\right.
=
\left.
\begin{array}{c|cccc}
c_1 & a_{11} & a_{12} & \cdots & a_{1v} \\
c_2 & a_{21} & a_{22} & \cdots & a_{2v} \\
\vdots & \vdots & \vdots & \ddots & \vdots \\
c_v & a_{v1} & a_{v2} & \cdots & a_{vv} \\
\hline
 & b_1 & b_2 & \cdots & b_v
\end{array}
\right.
\end{equation}

The Butcher tableau elegantly encompasses both explicit Runge--Kutta (ERK) method and implicit Runge--Kutta (IRK) method. For the ERK scheme, the matrix $A$ is strictly lower triangular (i.e., $a_{j,i}=0$ for $i\geq j$). The intermediate state approximation $\mathbf{\xi}_j$ is thus explicitly formulated as:

\begin{equation}\label{eq_ERK}
\begin{aligned}
\mathbf{\xi}_j &= \mathbf{u}_n + h \sum_{i=1}^{j-1} a_{j,i}\, f(t_n + c_i h, \mathbf{\xi}_i).
\end{aligned}
\end{equation}

This strict forward-dependency allows the stages to be evaluated sequentially. Consequently, exactly $\nu$ independent function evaluations are required, yielding a linear time complexity of $O(\nu)$ per step.

In the IRK method, $A$ is a dense matrix, which couples the intermediate stages $\mathbf{\xi}_1,\mathbf{\xi}_2,\dots,\mathbf{\xi}_{\nu}$, as illustrated in \eqref{eq_runge_kutta}. Typically, resolving these implicit equations involves computationally demanding nonlinear iterative solvers. However, within a specific switching interval, the buck converter operates as a Linear Time-Invariant (LTI) system, described by $f(t,\mathbf{u})=\mathbf{M}_S\mathbf{u}+\mathbf{d}_S$. This linearity allows the coupled stages to be directly analytically solved via Gaussian elimination as a single linear system. Since the computational cost of Gaussian elimination scales cubically with the matrix dimension \cite{golub2013matrix}, the time complexity for the IRK scheme becomes $O({\nu^3})$.

\subsection {Differentiable Physics Simulation for Buck Converter}
The aforementioned numerical integration schemes readily enable the forward simulation of the converter's dynamic states, such as $i_L$ and $v_o$. However, parameter estimation essentially solves the inverse problem. It aims to iteratively calibrate and optimize the physical parameters of interest by minimizing the discrepancy between the model-derived dynamic behaviors and the actual observations. To enable gradient-based optimization for this task, the numerical simulation must be fully differentiable with respect to the target parameters.

In the DP simulation framework, the time-stepping process is reformulated from the perspective of differentiable operator composition. Rather than viewing the numerical integration as a conventional black-box solver, each discrete calculation step is explicitly treated as a differentiable mathematical operator. The single time-stepping scheme derived in Section II-B is reformulated as:

\begin{equation}\label{eq_dpo1}
\begin{aligned}
\mathbf{u}_{n+1} &= \mathbf{u}_n + h \sum_{j=1}^{\nu} b_j\, f(t_n + c_j h, \mathbf{\xi}_j) =\mathcal{P}_{S,h}(\mathbf{u}_n,\boldsymbol{\theta}),
\end{aligned}
\end{equation}

\noindent
where $\mathcal{P}(\cdot)$ represents the functional operator encompassing the fundamental algebraic operations within the numerical method. $\boldsymbol{\theta}$ denotes the vector of target parameters to be estimated (i.e., ${\boldsymbol{\theta}=[ R_{dson}, v_F, L, R_L, C, R_C, R]^T}$). The state evolution over a given time interval can be mathematically formulated as a successive composition of these functional operators: 

\begin{equation}\label{eq_dpo2}
\begin{aligned}
\mathbf{u}_{n+q h_1}
=
\underbrace{
\mathcal{P}_{q}\circ \mathcal{P}_{q-1}\cdots \circ \mathcal{P}_{2} \circ \mathcal{P}_{1}
}_{q\ \text{steps}}
(\mathbf{u}_n,\boldsymbol{\theta}),
\end{aligned}
\end{equation}

\noindent
where $\circ$ denotes function composition, i.e., $f(g(x))=f \circ g(x)$. Let a given macroscopic time interval $\Delta t$ be uniformly divided into $q$ discrete steps with a step size of $h_1 = \Delta t / q$. Eq.\eqref{eq_dpo2} indicates that the final state vector $\mathbf{u}(t_n+\Delta t)=\mathbf{u}_{n+qh_1}$ is obtained by advancing the initial state $\mathbf{u}(t_n)=\mathbf{u}_{n}$ through $q$ successive discrete steps. It should be noted that to maintain the integrity of the gradient chain, every operator $\mathcal{P}_i(\cdot)$ must provide exact gradients with respect to both its inputs, namely $\partial \mathcal{P}_i / \partial \boldsymbol{\theta}$ and $\partial \mathcal{P}_i / \partial \mathbf{u}$. The implementation of automatic differentiation is described in detail in Section II-D.

Likewise, the DP simulation framework can be directly extended to power electronic converters. In this paper, the framework is applied to capture the dynamic relationship between the sparsely sampled data points of the buck converter, denoted as the state vector $\mathbf{u}_n = [i_{L(n)}, v_{o(n)}]^T$. These sampled data points serve as the foundation for model training, and the specific strategies for their selection and configuration will be elaborated in Section II-E.

However, the inherent switching actions of semiconductor devices periodically alter the circuit topology. The converter's dynamic behavior constitutes a non-smooth process that must be described by two distinct sets of ordinary differential equations (ODEs), corresponding to the $S=1$ and $S=0$ states established in \eqref{state_space_eq}. Conventionally, the extreme non-linearity induced by such abrupt topological transitions renders the global dynamic behavior non-differentiable at the switching nodes. Crucially, for the parameter inference task, it is only strictly required to maintain the continuity of the computational graph with respect to the target physical parameters. By treating the exact switching instant merely as a transition node where the mathematical operators are swapped, the uninterrupted propagation of gradients can be successfully maintained.

\begin{figure}
  \begin{center}
  \includegraphics[width=3.5in]{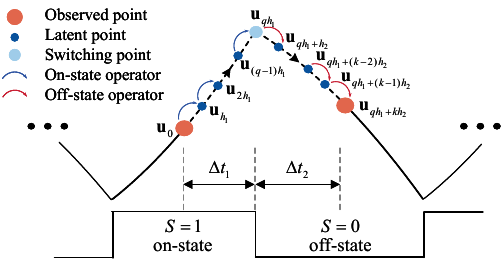}\\
  \caption{Numerical time-stepping scheme for the dynamic evolution between two sampling instants. The interval $\Delta t_1$ is propagated over $q$ steps and $\Delta t_2$ over $k$ steps, with operator switching performed at the switching node.}\label{DP_1}
  \end{center}
\end{figure}

Specifically, considering practical buck converter control designs, sampling points are typically located at the specific instants within a given switching state (e.g., the valley or peak of the triangle carrier), as illustrated in Fig.~\ref{DP_1}. Consequently, both the sampling instants and the switching nodes are deterministically known and readily accessible from the digital controller. Let $\Delta t_1$ and $\Delta t_2$ denote the exact time intervals between the observation points and the intermediate switching node. The dynamic evolution during $\Delta t_1$ is simulated using $q$ discrete steps with a micro-step size of $h_1 = \Delta t_1 / q$. Similarly, the subsequent interval $\Delta t_2$ is simulated over $k$ steps with a step size of $h_2 = \Delta t_2 / k$. Therefore, by seamlessly switching the numerical operators exactly at the switching node, the converter's piecewise-smooth dynamic behavior can be accurately captured as a differentiable function:

\begin{equation}\label{eq_dpo3}
\mathbf{u}_{qh_1 + kh_2} = \underbrace{\mathcal{P}_{off} \circ \dots  \circ \mathcal{P}_{off}}_{k\text{ steps}}\circ\underbrace{\mathcal{P}_{on} \circ \dots \circ \mathcal{P}_{on} }_{q\text{ steps}}(\mathbf{u}_0, \theta),
\end{equation}

\noindent
where $\mathcal{P}_{on}$ and $\mathcal{P}_{off}$ represent the numerical time-stepping operators corresponding to the ODEs of the on-state and off-state topologies, respectively. As a result, the physical parameters of the dc--dc converter can be effectively identified using gradient-based optimization algorithms within this architecture. Ultimately, the objective loss function under the proposed DP simulation framework is formulated as the Mean Squared Error (MSE) between the measured data and the model-derived states:

\begin{equation}\label{eq_dpo3}
\begin{split}
\mathcal{J}(\boldsymbol{\theta}) &= \frac{1}{m-1}\sum_{j=1}^{m-1} \left\| \mathbf{u}^{(j+1)}-\hat{\mathbf{u}}^{(j+1)} \right\|_2^2  \\
&= \frac{1}{m-1}\sum_{j=1}^{m-1} \bigl[ (i_L^{(j+1)} - \hat{i}_L^{(j+1)})^2 + (v_o^{(j+1)} - \hat{v}_o^{(j+1)})^2 \bigr]
\end{split}
\end{equation}

\noindent
where $j$ denotes the index of the sample, $m$ is the total number of data samples, $\mathbf{u}^{(j)}$ represents the actual observational data, and $\hat{\mathbf{u}}^{(j)}$ represents the corresponding simulated states. This framework design seamlessly bridges classical numerical integration methods with the rapid advancements in machine learning of the past two decades. Owing to its profound physical interpretability, the framework naturally guarantees that the inferred target parameters strictly conform to the underlying dynamic constraints of the converter.

\subsection {Automatic Differentiation and Backpropagation}

In the DP simulation framework, the dynamic behavior of the converter is mathematically modeled as a deep composition of functional operators. Directly deriving the exact symbolic derivatives of such a complex composition would incur prohibitive memory and computational overhead.
Automatic Differentiation (AD) is employed to efficiently extract the gradients of the target parameters by systematically applying the chain rule to the fundamental arithmetic operations comprising the dynamic simulation. This is achieved by complementing each intermediate variable $w_i$ in the computational graph with an adjoint variable:

\begin{equation}
\bar{w}_i = \frac{\partial y}{\partial w_i} \label{eq:adjoint},
\end{equation}

\begin{figure}
  \begin{center}
  \includegraphics[width=3.5in]{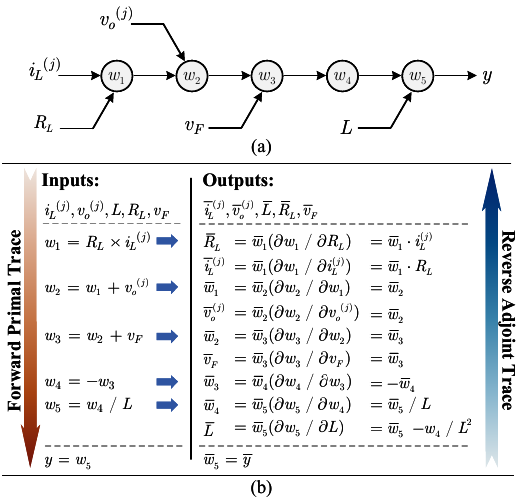}\\
  \caption{Derivative computation by reverse mode AD for the example $y= -({R_L i_L^{(j)} + v_o^{(j)} + v_F})/{L}$. (a) Computational graph with the parameters of interest taken as inputs; (b) Forward evaluation of the primal variables and reverse propagation of the corresponding adjoint variables.}\label{DP_2}
  \end{center}
\end{figure}

\noindent
which represents the sensitivity of a considered output $y$ with respect to changes in $w_i$. The derivative computation in AD fundamentally operates in two distinct phases. In the reverse mode AD\footnote{Automatic differentiation can be implemented in either forward or reverse mode, depending on how derivatives are propagated through the computational graph~\cite{baydin2018automatic}. Reverse mode is more efficient for evaluating the gradient of a scalar-valued objective with respect to a large number of parameters.}, derivative evaluation includes a forward phase and a reverse phase. Consider the calculation of the inductor current slope as an example, which is an elementary operation in converter dynamic simulation. When the MOSFET of the buck converter is in off-state (i.e., $S=0$), the inductor current slope is given by:

\begin{equation}
y=\frac{di_L^{(j)}}{dt} = -\frac{R_L i_L^{(j)} + v_o^{(j)} + v_F}{L}.\label{eq:i_Lslope}
\end{equation}

The corresponding derivative computation process for the intermediate variables and target parameters is shown in Fig.~\ref{DP_2}. In the forward phase, the function is evaluated and the dependency relations in the computational graph are recorded. In the backward phase, the adjoint variables are propagated from the output back to the inputs, and the required derivatives are obtained via the chain rule.

However, the output of a dynamic system is governed not only by its immediate preceding state but cumulatively by the entire sequence of past intermediate states.  Concurrently, as previously established, each elementary differentiable operator within the converter's dynamic simulation is parameterized by the same set of shared target parameters $\boldsymbol{\theta}$. To precisely apportion this temporal causal responsibility, Backpropagation Through Time (BPTT) is introduced as a powerful analytical tool~\cite{werbos2002backpropagation}. 

\begin{figure}
  \begin{center}
  \includegraphics[width=3.5in]{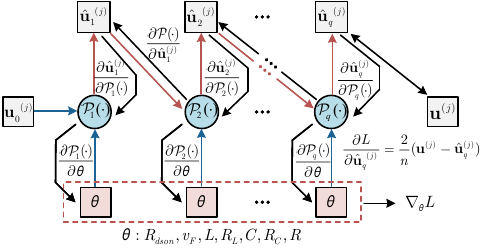}\\
  \caption{BPTT over multiple time steps for gradient computation with shared parameters $\boldsymbol{\theta}$.}\label{BPTT}
  \end{center}
\end{figure}

As illustrated in Fig.~\ref{BPTT}, BPTT unrolls the dynamic system along the time axis into a  differentiable computational graph. By backpropagating the error from the ultimate output, the derivatives with respect to the parameters at each discrete time step are evaluated. Accumulating these step-wise contributions to the shared parameters yields the total gradient of the objective function, which can be formulated as:

\begin{equation}\label{eq_gradient}
\begin{split}
\nabla_{\theta}\mathcal{J}
&=
\frac{2}{m-1}
\sum_{j=1}^{m-1}
\left(
\frac{\partial \hat{\mathbf{u}}_{q}^{(j+1)}}{\partial \theta}
\right)^{T}
\left(
\hat{\mathbf{u}}_{q}^{(j+1)} - \mathbf{u}^{(j+1)}
\right),
\\
\frac{\partial \hat{\mathbf{u}}_{q}^{(j+1)}}{\partial \theta}
&=
\frac{\partial \mathcal{P}_q}{\partial \hat{\mathbf{u}}_{q-1}^{(j+1)}}
\frac{\partial \mathcal{P}_{q-1}}{\partial \hat{\mathbf{u}}_{q-2}^{(j+1)}}
\cdots
\frac{\partial \mathcal{P}_2}{\partial \hat{\mathbf{u}}_{1}^{(j+1)}}
\frac{\partial \mathcal{P}_1}{\partial \theta}
+\cdots+
\frac{\partial \mathcal{P}_q}{\partial \theta}
\\
&=\sum_{\ell=1}^{q}
\left(
\prod_{r=q}^{\ell+1}
\frac{\partial \mathcal{P}_r}{\partial \hat{\mathbf{u}}_{r-1}^{(j+1)}}
\right)
\frac{\partial \mathcal{P}_{\ell}}{\partial \boldsymbol{\theta}}.
\end{split}
\end{equation}

This formulation successfully establishes the gradient flow for the target parameters within the DP simulation framework. This enables the direct application of advanced gradient-based optimization algorithms. Furthermore, it supports the efficient computation of higher-order derivatives, such as the Hessian matrix, thereby facilitating higher-order optimization strategies.

\subsection {Sampling Strategy and Data Configuration}

The aforementioned DP simulation framework is designed to numerically simulate the dynamic evolution between two sampled instants. Based on this simulated trajectory, the physical parameters that best match the observed responses can be iteratively learned. The selection of the sampling strategy is of central importance, as it affects the informativeness of the sampled data and the practical feasibility of the proposed method. 

\begin{figure}
  \begin{center}
  \includegraphics[width=3.5in]{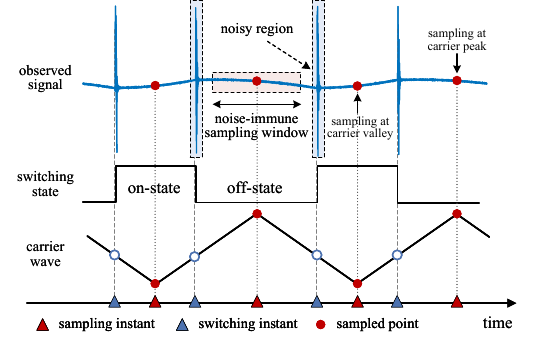}\\
  \caption{Carrier-synchronized sampling strategy for noise-immune data acquisition. The observed signal is sampled at predefined carrier valley and peak instants, while the switching ringing regions are excluded.}\label{data_sampling}
  \end{center}
\end{figure}

To maximize information yield, the sampling window is allocated to the transient intervals of the converter. Compared to steady state operation, transient dynamics are significantly more sensitive to parameter variations, thereby providing richer dynamic signatures for the estimation algorithm. In addition, Low-frequency sampling is highly preferred for compatibility with standard digital control systems. While peak-to-peak sampling captures the maximum signal variation within a switching cycle, its measurement accuracy is severely compromised by switching ringing. Consequently, carrier-synchronized sampling (i.e., capturing data exactly at the carrier valley and peak) emerges as a superior and more practical alternative. As illustrated in Fig.~\ref{data_sampling}, these predefined sampling instants are positioned away from the switching events, ensuring a noise-immune sampling environment. 

However, this sparse sampling strategy inevitably encompasses a topological switching event between adjacent observation points. Remarkably, the proposed framework effectively resolves this by leveraging the deterministic timing signals from the controller. Consequently, this unique cross-topology differentiable simulation drastically enhances the utilization of sampled data. 


\begin{figure}
  \begin{center}
  \includegraphics[width=3.5in]{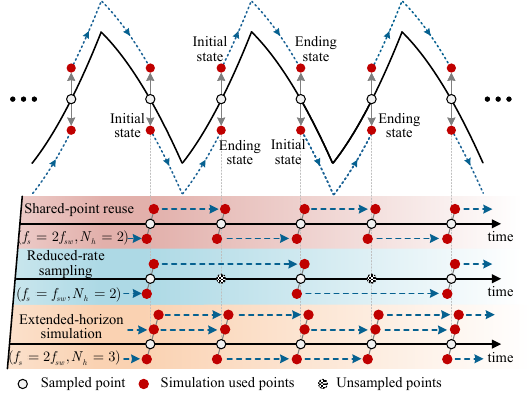}\\
  \caption{Potential data reuse strategies enabled by the proposed DP simulation framework, including shared-point reuse, reduced-rate sampling, and extended-horizon simulation.}\label{Data_Config}
  \end{center}
\end{figure}

As shown in Fig.~\ref{Data_Config}, each sampled point can seamlessly serve as both the ending state of the preceding interval and the initial state of the subsequent one. Likewise, the framework can be extended to a longer time scale, rather than being limited to the dynamic evolution between only two adjacent points. In other words, a single sampled point can be used to infer the subsequent $N_h-1$ sampled points, thereby enabling the learning of the dynamic behavior over a time window of length $N_h$. This also indicates the potential to further reduce the sampling frequency. Specifically, this capability allows the sampling frequency to be reduced from $2f_{sw}$ to $f_{sw}$, or even lower to one sample over multiple switching cycles.

These configurations highlight the flexibility of the proposed DP simulation framework in organizing and reusing sampled data. In the subsequent study, the basic one-step prediction setting is considered first. The effects of additional formulations, including an extended-horizon simulation and other constraints, are then further examined.

\section{Simulation Verification and Robustness analysis}
To verify the accuracy and robustness of the proposed framework, a dc-–dc buck converter is implemented in MATLAB/Simulink to generate simulation data. Multiple converter configurations with different operating conditions and hardware parameters are considered to evaluate the generalization capability of the method, as listed in Table \ref{tab:cases_config}. Notably, the proposed framework offers highly flexible structural designs and data utilization. The tradeoff between estimation accuracy and computational complexity is also evaluated. Furthermore, uncertainty may be introduced at different stages in practical condition monitoring. Thus, both the aleatoric uncertainty (e.g., measurement noise, ADC quantization error, and synchronization offset) and epistemic uncertainty (e.g., model--plant mismatch) are further introduced into the clean simulation data for robustness analysis.

\begin{table}[htbp]
\centering
\caption{Representative Buck Converter Configurations for Generalization Tests}
\label{tab:cases_config}
\begin{threeparttable}
\setlength{\tabcolsep}{4pt}
\begin{tabular}{c c l}
\toprule
Case & Type & Configuration \\
\midrule
I & Ref. & \makecell[l]{$V_{\mathrm{in}}/V_{\mathrm{ref}} = 48~\mathrm{V}/24~\mathrm{V}$, $f_{\mathrm{sw}} = 20~\mathrm{kHz}$,\\
$L/R_L = 725~\mu\mathrm{H}/0.314~\Omega$, $R_{\mathrm{DS(on)}} = 0.151~\Omega$,\\
$C/R_C = 164.5~\mu\mathrm{F}/0.201~\Omega$, $V_F = 1~\mathrm{V}$,\\$R_0/R_1/R_2/R_3 = 6.2~\Omega/10.2~\Omega/3.1~\Omega/6.1~\Omega$}
\\
\arrayrulecolor{gray!35}
\specialrule{0.3pt}{0pt}{0.5ex}
\arrayrulecolor{black}
II & OpCond. &
Same as Case I except: $V_{\mathrm{in}}/V_{\mathrm{ref}} = 48~\mathrm{V}/12~\mathrm{V}$ \\
\arrayrulecolor{gray!35}
\specialrule{0.3pt}{0pt}{0.5ex}
\arrayrulecolor{black}
III & OpCond. &
Same as Case I except: $f_{\mathrm{sw}} = 100~\mathrm{kHz}$ \\
\arrayrulecolor{gray!35}
\specialrule{0.3pt}{0pt}{0.5ex}
\arrayrulecolor{black}
IV & OpCond. & \makecell[l]{Same as Case I except:\\
$R_0/R_1/R_2/R_3 = 6.2~\Omega/1.2~\Omega/3.8~\Omega/1.4~\Omega$}\\
\arrayrulecolor{gray!35}
\specialrule{0.3pt}{0pt}{0.5ex}
\arrayrulecolor{black}
V & HwParam. &\makecell[l]{Same as Case I except:\\
$L/R_L = 278.5~\mu\mathrm{H}/0.254~\Omega$,
$R_{\mathrm{DS(on)}} = 0.082~\Omega$,\\
$C/R_C = 124.5~\mu\mathrm{F}/0.101~\Omega$,
$V_F = 0.8~\mathrm{V}$ }
\\
\arrayrulecolor{gray!35}
\specialrule{0.3pt}{0pt}{0.5ex}
\arrayrulecolor{black}
VI & OpCond. &\makecell[l]{Same as Case V except:\\
$R_0/R_1/R_2/R_3 = 6.2~\Omega/1.2~\Omega/3.8~\Omega/1.4~\Omega$}\\
\arrayrulecolor{gray!35}
\specialrule{0.3pt}{0pt}{0.5ex}
\arrayrulecolor{black}
VII & HwParam. &\makecell[l]{Same as Case I except:\\
$L/R_L = 278.5~\mu\mathrm{H}/0.554~\Omega$,
$R_{\mathrm{DS(on)}} = 0.282~\Omega$,\\
$C/R_C = 124.5~\mu\mathrm{F}/0.351~\Omega$,
$V_F = 1.5~\mathrm{V}$ }\\
\arrayrulecolor{gray!35}
\specialrule{0.3pt}{0pt}{0.5ex}
\arrayrulecolor{black}
VIII & OpCond. &\makecell[l]{Same as Case VII except:\\
$R_0/R_1/R_2/R_3 = 6.2~\Omega/1.2~\Omega/3.8~\Omega/1.4~\Omega$}\\
\bottomrule
\end{tabular}
\begin{tablenotes}[flushleft]
\footnotesize
\item Ref.: reference case; OpCond.: cases with modified operating conditions; HwParam.: cases with modified hardware parameters.
\end{tablenotes}
\end{threeparttable}
\end{table}

\subsection {Accuracy and Computation Complexity}
\begin{figure}
  \begin{center}
  \includegraphics[width=3.5in]{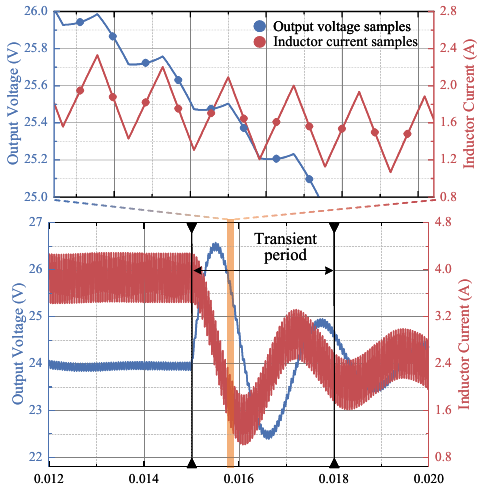}\\
  \caption{Carrier-synchronized acquisition of inductor current and output voltage samples during a converter transient.}\label{transient}
  \end{center}
\end{figure}

Training data are collected by carrier-synchronized sampling of the inductor current and output voltage during converter transients, as shown in Fig. \ref{transient}. Such transient processes are common in practical converters. They occur during startup and shutdown, load switching, and control mode transition. Moreover, the required observations are also part of the feedback signals used in the control loop. Therefore, the data required by the proposed method are readily accessible in practical applications.

However, as the switching frequency increases and control algorithms continue to improve, the duration of the transient process becomes shorter. This makes data collection more challenging. It should be noted that part of the steady-state data can also be included in the training set to partially alleviate data scarcity. More importantly, the degradation of power electronic components in a well-designed converter is typically a very slow process (e.g., over months or even years). This provides sufficient time for data collection, especially considering the low data demand of the proposed method with strong physical interpretability.

\begin{table}[htbp]
\centering
\caption{Optimizer Hyperparameters Used in The Two-stage Training Procedure}
\label{tab:optimizer_para_setting}
\begin{tabular}{lll}
\hline
Optimizer & Parameter & Value \\
\hline
\multirow{4}{*}{\begin{tabular}[c]{@{}c@{}}Adam\\Method\end{tabular}}
& Learning rate $lr$ & $0.001$ \\
& First-moment decay rate $\beta_1$ & $0.9$ \\
& Second-moment decay rate $\beta_2$ & $0.999$ \\
& number of epochs & 2000\\
\hline
\multirow{7}{*}{\begin{tabular}[c]{@{}c@{}}L-BFGS\\Method\end{tabular}}
& Learning rate $lr$ & $1$ \\
& Maximal number of iterations & 5000 \\
& Maximal number of function evaluations & 50000 \\
& Gradient tolerance & $10^{-10}$ \\
& Change tolerance & $10^{-12}$ \\
& History size & $50$ \\
& Line search method  & strong wolfe \\
\hline
\end{tabular}
\end{table}

\begin{table*}[t]
\caption{Percentage Errors (\%) and Computational Cost under Different Time-Stepping Strategies}
\label{tab:rk_config}
\centering
\begin{threeparttable}
\footnotesize

\begin{tabular}{l | c c c c c c c c c c | l | l | l}
\hline
\diagbox[dir=NW]{\makecell[c]{Methods}}{\makecell[c]{Error (\%)}}
& $L$ & $R_L$ & $C$ & $R_C$ & $R_{dson}$ & $V_F$ & $R_D$ & $R_1$ & $R_2$ & $R_3$ &  \makecell{Fwd\\FLOPs} & \makecell{Avg.\\iter time} & \makecell{Obs.\\MSE} \\
\hline
ERK1 (substeps=1) & 0.52 & 18.45 & 0.43 & 23.26 & 40.71 & 18.21 & 0.76 & 0.41 & 0.40 & 0.46 & \makecell[l]{$\approx{114}$} & 5.76 ms & 5.94e-5\\
ERK1 (substeps=5) & 0.22 & 10.34 & 0.20 & 7.12 & 26.94 & 7.76 & 1.77 & 0.11 & 0.06 & 0.16 &  \makecell[l]{$\approx{354}$} & 9.94 ms & 2.45e-5\\
ERK1 (substeps=10) & 0.19 & 9.36 & 0.17 & 5.11 & 25.36 & 6.52 & 1.90 & 0.07 & 0.02 & 0.12 & \makecell[l]{$\approx{654}$} & 15.09 ms & 2.33e-5\\
ERK2 (substeps=1) & 0.18 & 8.70 & 0.12 & 2.91 & 24.78 & 5.23 & 2.17 & 0.02 & 0.03 & 0.08 & \makecell[l]{$\approx{218}$} & 5.91 ms & 2.28e-5\\
ERK2 (substeps=5) & 0.15 & 8.18 & 0.15 & 3.10 & 23.01 & 5.23 & 1.95 & 0.03 & 0.02 & 0.09 &  \makecell[l]{$\approx{874}$} & 10.46 ms & 2.28e-5\\
ERK2 (substeps=10) & 0.15 & 8.39 & 0.15 & 3.11 & 23.77 & 5.29 & 2.05 & 0.03 & 0.02 & 0.09 & \makecell[l]{$\approx{1694}$} & 16.03 ms & 2.28e-5\\
ERK4 (substeps=1) & 0.15 & 8.29 & 0.15 & 3.11 & 23.39 & 5.26 & 2.00 & 0.03 & 0.02 & 0.09 &  \makecell[l]{$\approx{618}$} & 5.89 ms & 2.28e-5\\
ERK4 (substeps=5) & 0.15 & 8.35 & 0.15 & 3.11 & 23.60 & 5.28 & 2.03 & 0.03 & 0.02 & 0.09 &  \makecell[l]{$\approx{2874}$} & 10.61 ms & 2.28e-5\\
ERK4 (substeps=10) & 0.15 & 8.35 & 0.15 & 3.11 & 23.60 & 5.28 & 2.03 & 0.03 & 0.02 & 0.09 &  \makecell[l]{$\approx{5694}$} & 16.13 ms & 2.28e-5\\
ERK10 (substeps=1) & 0.15 & 8.35 & 0.15 & 3.11 & 23.60 & 5.28 & 2.03 & 0.03 & 0.02 & 0.09 &  \makecell[l]{$\approx{2186}$} & 6.05 ms & 2.28e-5\\
ERK10 (substeps=5) & 0.15 & 8.36 & 0.15 & 3.11 & 23.63 & 5.28 & 2.03 & 0.03 & 0.02 & 0.09 &  \makecell[l]{$\approx{10714}$} & 10.66 ms & 2.28e-5\\
ERK10 (substeps=10) & 0.15 & 8.35 & 0.15 & 3.11 & 23.60 & 5.28 & 2.03 & 0.03 & 0.02 & 0.09 &  \makecell[l]{$\approx{21374}$} & 16.23 ms & 2.28e-5\\
IRK2 (substeps=1) & 0.15 & 8.39 & 0.15 & 3.11 & 23.73 & 5.29 & 2.04 & 0.03 & 0.02 & 0.09 &  \makecell[l]{$\approx{336}$} & 6.15 ms & 2.28e-5\\
IRK4 (substeps=1) & 0.15 & 8.35 & 0.15 & 3.11 & 23.60 & 5.28 & 2.03 & 0.03 & 0.02 & 0.09 & \makecell[l]{$\approx{1429}$} & 6.11 ms & 2.28e-5\\
IRK10 (substeps=1) & 0.15 & 8.35 & 0.15 & 3.11 & 23.60 & 5.28 & 2.03 & 0.03 & 0.02 & 0.09 & \makecell[l]{$\approx{14821}$} & 6.55 ms & 2.28e-5\\
IRK20 (substeps=1) & 0.15 & 8.35 & 0.15 & 3.11 & 23.60 & 5.28 & 2.03 & 0.03 & 0.02 & 0.09 & \makecell[l]{$\approx{101567}$} & 7.61 ms & 2.28e-5\\
\hline
\end{tabular}
\begin{tablenotes}[flushleft]
\footnotesize
\item Fwd FLOPs: Approximate floating-point operations for one forward pass of a single sample; Avg iter time: Average wall-clock time per Adam iteration, measured on an NVIDIA GeForce RTX 5060 Ti; Obs. MSE: Mean squared error between the reconstructed and measured observations after training.
\end{tablenotes}
\end{threeparttable}
\end{table*}

In this simulation study, converter transients are triggered by load variations ($ R_0 \to R_1, \, R_1 \to R_2, \, R_2 \to R_3 $). For each transient case, the sampled data over 30 switching cycles are collected as the training data for one load transition, so as to match the response speed encountered in practical converters. The data from the three load transitions are then combined to form the training dataset ($D=\{(v_n,i_n,\delta t,s),( v_{n+1},i_{n+1})\}_m$), which is used for converter condition monitoring. The DP simulation framework is implemented in Python with PyTorch. Because the framework establishes a differentiable computational graph for converter dynamics, gradient-based optimizers can be directly employed for training. In this work, the Adam \cite{kingma2014adam} optimizer followed by a full-batch L-BFGS \cite{liu1989limited} optimizer is adopted, with the parameter settings listed in Table~\ref{tab:optimizer_para_setting}.

The parameter estimation process relies on forward simulation of the converter dynamics. Since the discrete formulation of the differentiable operator may deviate from the underlying continuous-time physical process, the order of the Runge--Kutta method and the number of integration steps per switching interval should be properly selected to keep the discretization error within an acceptable range. Taking Case I as an example, the parameter estimation performance under different Runge--Kutta configurations is first evaluated using clean simulation data. 
The results are summarized in Table~\ref{tab:rk_config}.

It can be observed that $C$, $L$, and $R_{\mathrm{load}}$ can be accurately identified. For clarity, the set of accurately identifiable parameters is defined as $\boldsymbol{\theta}_1=\{C,L,R_{\mathrm{load}}\}$. However, several other parameters, especially $R_L$ and $R_{dson}$, are difficult to estimate individually with high accuracy. This is because these two resistive components have nearly identical effects on the observed variables $v_o$ and $i_L$, which has also been discussed in detail in previous studies~\cite{Digital2021Pyz,Physics2025Osy}. 
Therefore, the equivalent resistance $R_D = R_L + R_{dson}$ is adopted for monitoring the degradation associated with the inductor winding resistance and the power semiconductor on-state resistance.

As shown in Table~\ref{tab:rk_config}, once the Runge--Kutta scheme reaches a sufficient order, further increasing the integration order or reducing the integration step size brings only marginal improvement in parameter identification. 
This is because the truncation error of a $p$th-order Runge--Kutta method scales with a high-order term of the integration step size. 
When each switching interval $\Delta t$ is divided into $N_{\mathrm{sub}}$ substeps, the substep size becomes $\Delta t/N_{\mathrm{sub}}$, and the corresponding error term can be qualitatively related to $O((\Delta t/N_{\mathrm{sub}})^p)$. 
In this case, the numerical integration error is already much smaller than the errors introduced by observation noise and model--plant mismatch in practical condition monitoring.

In addition, benefiting from GPU-based parallel computation, different Runge--Kutta orders lead to only minor differences in practical runtime. 
In contrast, the number of integration substeps has a more direct impact on the computational cost, because the proposed framework is inherently constrained by the sequential nature of time-marching simulation and gradient propagation. 
Considering estimation accuracy, computational cost, and applicability to practical converters, the single-step classical fourth-order Runge--Kutta method is selected for simulating the converter dynamics in the subsequent experiments. 
The parameter identification results under different converter configurations are reported in Table~\ref{tab:error_cases}.

\begin{table}[t]
\caption{Parameter Identification Performance (\%) Under Different Converter Configurations}
\label{tab:error_cases}
\centering
\footnotesize
\setlength{\tabcolsep}{5pt}

\begin{tabular}{c c c c c c c @{\hspace{1.7em}} c}
\toprule
Case & \makecell{$\hat{\boldsymbol{\theta}}_1$\\(\%)} & \makecell{$R_C$\\(\%)} &  \makecell{$V_F$\\(\%)} & \makecell{$R_L$\\(\%)} & \makecell{$R_{dson}$\\(\%)} &\makecell{$R_D$\\(\%)}  & \makecell{Obs.\\MSE} \\
\midrule
I    & 0.10 & 3.11  & 5.26  & 8.23  & 23.39  & 2.00  & 2.28e-5 \\
II   & 0.10 & 2.45  & 0.10  & 14.42 & 95.29  & 21.21 & 2.26e-5 \\
III  & 0.10 & 0.10  & 0.10  & 0.46  & 1.45   & 0.16  & 2.17e-7 \\
IV   & 0.10 & 0.75  & 1.60  & 0.57  & 1.00   & 0.10  & 2.20e-5 \\
V    & 0.31 & 30.27 & 17.93 & 21.24 & 87.29  & 5.25  & 1.53e-4 \\
VI   & 0.10 & 6.72  & 11.60 & 14.68 & 70.35  & 6.07  & 1.43e-4 \\
VII  & 1.50 & 24.76 & 7.86  & 31.99 & 100.00 & 12.53 & 1.67e-4 \\
VIII & 0.25 & 2.25  & 19.27 & 14.70 & 34.58  & 1.92  & 1.47e-4 \\
\bottomrule
\end{tabular}
\vspace{0.4em}

\parbox{0.92\linewidth}{\footnotesize
$\hat{\boldsymbol{\theta}}_1=\{L,C,R_1,R_2,R_3\}$; its reported error is the mean of these parameter errors. Values below $0.1\%$ are rounded up to $0.1\%$ for clarity.
}
\end{table}

\subsection {Regularization Constraint}
As discussed above, several parameters remain difficult to identify accurately, even when the MSE of the training loss has been reduced to the order of $10^{-5}$. This indicates that a very small mismatch in the simulated dynamic waveform may correspond to a large deviation in certain parameter values. Such behavior is typical of an ill-posed inverse problem.

A straightforward remedy is to constrain the weakly identifiable parameters within a narrow prior range and use this range as a regularization condition. However, in application-oriented scenarios, the device parameters of a converter may vary over a wide range throughout its lifetime. It is hard to justify and standardize such prior bounds. Therefore, it is more reasonable to impose additional constraints from the converter dynamics themselves. In this work, two types of constraints are introduced: bidirectional consistency and long-horizon consistency. 

\begin{figure}
  \begin{center}
  \includegraphics[width=3.5in]{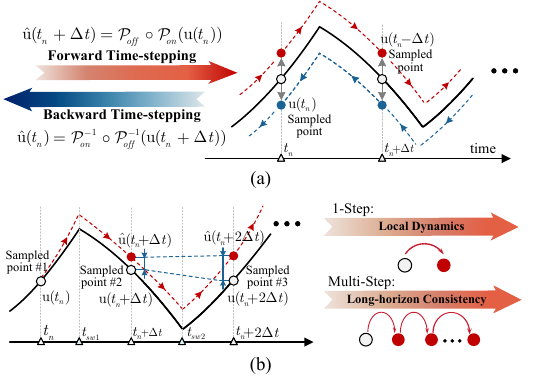}\\
  \caption{Illustration of the proposed regularization constraints. (a) Bidirectional consistency based on forward and backward time-stepping over the sampling interval. (b) Long-horizon consistency based on matching the measured trajectory over an extended prediction window.}\label{Regularization}
  \end{center}
\end{figure}

\begin{enumerate}
\item \textit{Bidirectional consistency:}
The physical operation of a converter evolves in the forward time direction. However, its dynamic behaviors are described by differential equations. From the viewpoint of numerical simulation, a valid parameter set should be locally consistent in both time directions. As illustrated in Fig.~\ref{Regularization} (a), the model should predict the next state from the current state and also reconstruct the current state from the next state over a short sampling interval. Accordingly, the loss function is defined as:

\begin{equation}\label{bi_constraint}
\begin{aligned}
\mathcal{J}_{bi}(\boldsymbol{\theta}) = \sum_{j=1}^{m-1} 
\left(
\frac{ \left\| \mathbf{u}^{(j+1)} - \hat{\mathbf{u}}_+^{(j+1)} \right\|_2^2 + \left\| \mathbf{u}^{(j)} - \hat{\mathbf{u}}_-^{(j)} \right\|_2^2 }{2(m-1)}
\right)
\end{aligned}
\end{equation}

\noindent
where $\hat{\mathbf{u}}_+^{(j+1)}$ is obtained by forward time-stepping from $\mathbf{u}^{(j)}$, and $\hat{\mathbf{u}}_-^{(j)}$ is obtained by backward time-stepping from $\mathbf{u}^{(j+1)}$.

\item \textit{Long-horizon Consistency:}
The state-space equations in ~\eqref{state_space_eq} show that some parameters, such as $R_L$, $R_{dson}$ and $v_F$, have strongly coupled effects on the observable states (i.e., $i_L$ and $v_o$). Therefore, one-step prediction may still achieve a very small loss even when these parameters are not accurately identified. This is because the compensation among coupled parameters can reproduce the local converter dynamics.

Long-horizon consistency is introduced to reduce this compensation effect. Instead of matching only one-step transitions, the model is required to reproduce the measured trajectory over an extended window. If the estimated parameters are inaccurate, small modeling errors may accumulate over the prediction horizon and become more evident, as shown in Fig.~\ref{Regularization} (b). Therefore, multi-step prediction provides a stronger dynamic constraint and guides the optimizer toward more physically meaningful parameter estimates. Under the long-horizon consistency constraint, the loss function is derived as:

\begin{equation}
\label{long_constraint}
\mathcal{J}_{\mathrm{lh}}(\boldsymbol{\theta})
=
\frac{\sum_{j=1}^{m-N_h+1}
\sum_{k=1}^{N_h-1}
\left\|
\hat{\mathbf{u}}_{k}^{(j+k)}
-
\mathbf{u}_{k}^{(j+k)}
\right\|_2^2}
{(m-N_h+1)(N_h-1)}
\end{equation}

\noindent
where $N_h$ is the window length of extended horizon, and $k$ denotes the $k$-th simulated point.
\end{enumerate}

Based on the constraints introduced above, the proposed regularization strategy is further tested using transient simulation data from different buck converter configurations. As shown in Table~\ref{tab:regularization_cases}, the proposed constraints generally improve the identifiability of the estimated parameters. For clean simulation data, the estimation accuracy generally improves as the prediction window length increases. Even the strongly coupled parameters
$R_L$ and $R_{dson}$ become more distinguishable when a longer dynamic trajectory is included in the training loss. However, in cases with small parasitic resistances and relatively small load variations, such as Cases V and VI, the accurate estimation of $R_L$ and $R_{dson}$ remains challenging.

The selection of the prediction window length is related to both the amount of training data and the computational cost. A longer window requires more transient samples for training, and the computational time of differentiable converter simulation increases approximately linearly with the window length. Therefore, an excessively long window may lead to unnecessary training cost. When the window length is set to $N_h=16$, the training time of the differentiable physical model can be kept within one minutes on an NVIDIA GeForce RTX 5060 Ti. With this setting, $ \hat{\boldsymbol{\theta}}_1$, $R_C$, $V_F$ and $R_D$ achieve errors below $1\%$ in most cases. Therefore, considering both computational cost and estimation accuracy, $N_h=16$ is selected for the subsequent robustness analysis.

\begin{table}[t]
\caption{Parameter Identification Performance (\%) With Regularization Constraint}
\label{tab:regularization_cases}
\centering
\footnotesize
\setlength{\tabcolsep}{5pt}

\begin{tabular}{c l c c c c c c}
\hline
Case & Constraint & \makecell{$\hat{\boldsymbol{\theta}}_1$\\(\%)} & \makecell{$R_C$\\(\%)} &  \makecell{$V_F$\\(\%)} & \makecell{$R_L$\\(\%)} & \makecell{$R_{dson}$\\(\%)} &\makecell{$R_D$\\(\%)}\\
\hline
\multirow{6}{*}{\begin{tabular}[c]{@{}c@{}}I\end{tabular}} 
& None & 0.10 & 3.11 & 5.26 & 8.29 & 23.39 & 2.00 \\
& Bi, L=3  & 0.10 & 0.63 & 1.94 & 5.08 & 16.45 & 1.91 \\
& Bi, L=6  & 0.10 & 0.20 & 0.91 & 3.51 & 11.73 & 1.44 \\
& Bi, L=11 & 0.10 & 0.10 & 0.25 & 1.87 & 6.52 & 0.86 \\
& Bi, L=16 & 0.10 & 0.10 & 0.21 & 0.80 & 2.68 & 0.33 \\
& Bi, L=21 & 0.10 & 0.10 & 0.24 & 0.75 & 2.51 & 0.31 \\
\arrayrulecolor{gray!35}
\specialrule{0.3pt}{0pt}{0.5ex}
\arrayrulecolor{black}
\multirow{2}{*}{\begin{tabular}[c]{@{}c@{}}II\end{tabular}} 
& None & 0.10 & 2.45 & 0.10 & 14.42 & 95.29 & 21.20 \\
& Bi, L=16 & 0.10 & 0.30 & 0.36 & 3.39 & 20.86 & 4.48 \\
\arrayrulecolor{gray!35}
\specialrule{0.3pt}{0pt}{0.5ex}
\arrayrulecolor{black}
\multirow{2}{*}{\begin{tabular}[c]{@{}c@{}}III\end{tabular}} 
& None & 0.10 & 0.10 & 0.10 & 0.46 & 1.45 & 0.16 \\
& Bi, L=16 & 0.10 & 0.10 & 0.10 & 0.10 & 0.10 & 0.10 \\
\arrayrulecolor{gray!35}
\specialrule{0.3pt}{0pt}{0.5ex}
\arrayrulecolor{black}
\multirow{2}{*}{\begin{tabular}[c]{@{}c@{}}IV\end{tabular}} 
& None & 0.10 & 0.75 & 1.60 & 0.57 & 1.00 & 0.10 \\
& Bi, L=16 & 0.10 & 0.10 & 0.21 & 0.31 & 0.92 & 0.10 \\
\arrayrulecolor{gray!35}
\specialrule{0.3pt}{0pt}{0.5ex}
\arrayrulecolor{black}
\multirow{2}{*}{\begin{tabular}[c]{@{}c@{}}V\end{tabular}} 
& None & 0.31 & 30.27 & 17.93 & 21.24 & 87.29 & 5.25 \\
& Bi, L=16 & 0.10 & 0.74 & 1.24 & 17.82 & 100 & 10.93 \\
\arrayrulecolor{gray!35}
\specialrule{0.3pt}{0pt}{0.5ex}
\arrayrulecolor{black}
\multirow{2}{*}{\begin{tabular}[c]{@{}c@{}}VI\end{tabular}} 
& None & 0.10 & 6.72 & 11.60 & 14.68 & 70.35 & 6.07 \\
& Bi, L=16 & 0.10 & 0.10 & 3.10 & 15.51 & 79.74 & 7.74 \\
\arrayrulecolor{gray!35}
\specialrule{0.3pt}{0pt}{0.5ex}
\arrayrulecolor{black}
\multirow{2}{*}{\begin{tabular}[c]{@{}c@{}}VII\end{tabular}} 
& None & 1.50 & 24.76 & 7.86 & 31.99 & 100.00 & 12.53 \\
& Bi, L=16 & 0.12 & 0.37 & 0.11 & 3.61 & 11.85 & 1.60 \\
\arrayrulecolor{gray!35}
\specialrule{0.3pt}{0pt}{0.5ex}
\arrayrulecolor{black}
\multirow{2}{*}{\begin{tabular}[c]{@{}c@{}}VIII\end{tabular}} 
& None & 0.25 & 2.25 & 19.27 & 14.70 & 34.58 & 1.92 \\
& Bi, L=16 & 0.10 & 0.10 & 0.43 & 0.44 & 0.92 & 0.10 \\
\hline
\end{tabular}
\vspace{0.4em}

\parbox{0.92\linewidth}{\footnotesize
$\hat{\boldsymbol{\theta}}_1=\{L,C,R_1,R_2,R_3\}$; its reported error is the mean of these parameter errors. Values below $0.1\%$ are rounded up to $0.1\%$ for clarity.
}
\end{table}

\subsection {Robustness Analysis}

\begin{figure}
  \begin{center}
  \includegraphics[width=3.5in]{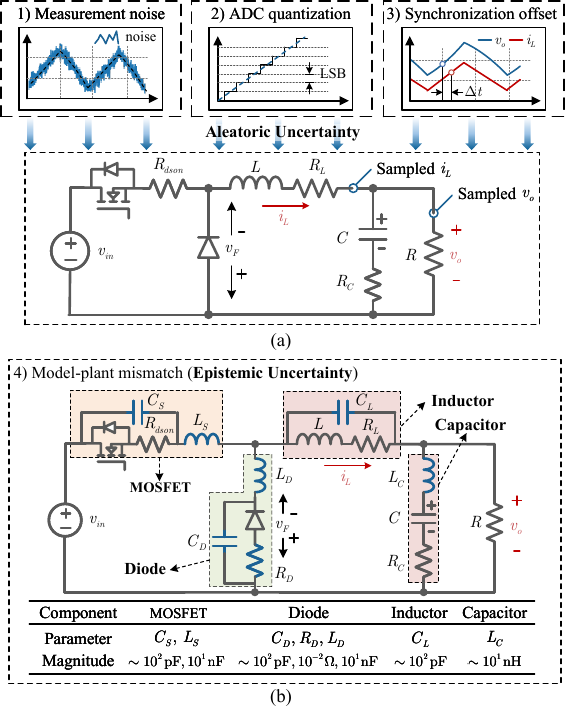}\\
  \caption{Uncertainty modeling for robustness evaluation. (a) Aleatoric uncertainty imposed on the sampled signals through measurement noise, ADC quantization, and synchronization offset. (b) Epistemic uncertainty introduced by model–-plant mismatch with added parasitic components.}\label{robustness_eval}
  \end{center}
\end{figure}

In practical operation and maintenance scenarios, the accuracy of converter parameter estimation can be affected by multiple sources of uncertainty. On the one hand, aleatoric uncertainty arises from the inherent randomness of the system, including sensor measurement noise, ADC quantization, and synchronization offset among multiple signals. On the other hand, epistemic uncertainty results from incomplete knowledge of the actual system \cite{10608153}, leading to model--plant mismatch between the simplified mathematical model and the real converter. These uncertainty factors are introduced into the clean simulation data to further evaluate the robustness of the proposed method, as illustrated in Fig.~\ref{robustness_eval}.

\begin{table*}[t]
\caption{Percentage Error (\%) of Component Parameters under Different Uncertainty Sources in Case I}
\label{tab:uncertainty_cases}
\centering
\footnotesize

\begin{tabular}{c | c c c c c c c c c c}
\hline
\diagbox[dir=NW]{\makecell[c]{Uncertainties}}{\makecell[c]{Error (\%)}}
& $L$ & $R_L$ & $C$ & $R_C$ & $R_{dson}$ & $V_F$ & $R_D$ & $R_1$ & $R_2$ & $R_3$  \\
\hline
Clean data & 0.10 & 0.80 & 0.10 & 0.10 & 2.68 & 0.21 & 0.33 & 0.10 & 0.10 & 0.10 \\
Model--plant mismatch         & 0.10 & 6.50 & 0.10 & 0.10 & 46.20 & 2.02 & 10.62 & 0.10 & 0.10 & 0.10 \\
ADC quantization error       & 0.10 & 2.91 & 0.10 & 0.28 & 9.78 & 0.36 & 1.21 & 0.10 & 0.10 & 0.10 \\
Synchronization offset (1 $\mu s$) & 0.10 & 6.41 & 0.10 & 1.10 & 20.45 & 3.12 & 2.31 & 0.10 & 0.10 & 0.10 \\
Measurement noise (5\textperthousand )   & 0.68 & 15.46 & 0.93 & 2.61 & $--$ & 0.65 & 7.87 & 0.15 & 0.10 & 0.27 \\
Measurement noise (1\%)      & 1.40 & 41.80 & 2.66 & 0.34 & $--$  & 7.09 & 18.21 & 0.81 & 0.59 & 0.45 \\
ADC-Sync-5\textperthousand noise    & 0.10 & 23.71 & 0.35 & 0.37 & $--$ & 2.45 & 11.41 & 0.13 & 0.30 & 0.10 \\
ADC-Sync-1\%noise      & 0.50 & 16.32 & 2.00 & 8.17 & $--$ & 4.12 & 6.57 & 0.10 & 0.25 & 0.54 \\
ADC-Sync-1\%noise-MPM       & 0.17 & 1.57 & 0.27 & 3.30 & 24.58 & 3.84 & 6.93 & 0.71 & 0.39 & 0.10 \\
\hline
\end{tabular}

\vspace{0.35em}
\footnotesize
\raggedright
(MPM: model--plant mismatch; ADC: ADC quantization error; Sync: synchronization offset). Values below $0.10\%$ are rounded up to $0.10\%$, and values above $50\%$ are represented by "$--$" .
\end{table*}

\begin{figure*}[!t]
  \begin{center}
  \includegraphics[width=\textwidth]{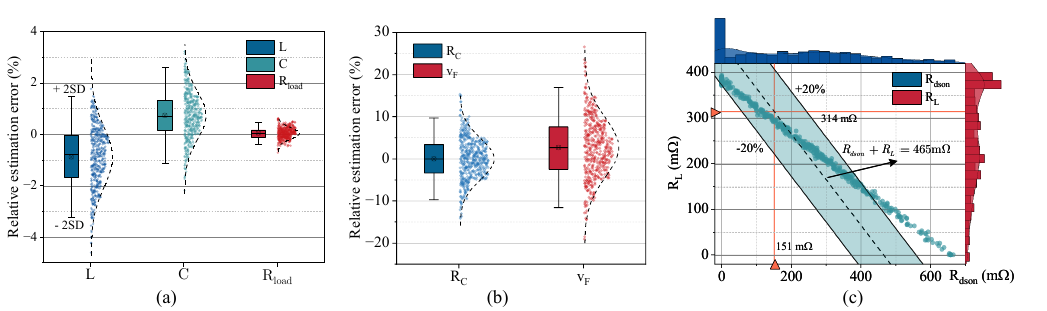}\\
  \caption{Parameter estimation results of Monte Carlo analysis in Case I. (a) Relative estimation errors of $L$, $C$, and $R_{\mathrm{load}}$. (b) Relative estimation errors of $R_C$ and $v_F$. (c) Joint distribution of $R_{dson}$ and $R_L$, where most samples fall within the shaded $\pm20\%$ band corresponding to the equivalent resistance $R_D=R_{dson}+R_L$.}\label{montecarlo}
  \end{center}
\end{figure*}

\begin{enumerate}
\item \textit{Measurement noise:}
Sensors and analog front-end circuits are subject to thermal noise, environmental disturbances, and variations in device properties. As a result, the measured signals inevitably contain random deviations. Zero-mean Gaussian noise is added to the sampled signals (e.g., $i_L$ and $v_o$) to emulate this uncertainty. The accuracy of a sampling system is usually specified with respect to the full-scale (FS) range. Specifically, considering an accuracy of $\pm0.5\%$ FS, the maximum error of output voltage $v_o$ is calculated as $30~\mathrm{V}\times0.5\%=150~\mathrm{mV}$, and that of inductor current $i_L$ is $25~\mathrm{A}\times0.5\%=125~\mathrm{mA}$. According to the $3\sigma$ rule of the Gaussian distribution, the standard deviations(i.e., $\sigma$) of the voltage and current noise are set to $50~\mathrm{mV}$ and $41.67~\mathrm{mA}$ respectively. Similarly, for $\pm1\%$ FS accuracy, the standard deviations are set to $\sigma_{v_o}=100~\mathrm{mV}$ and $\sigma_{i_L}=83.33~\mathrm{mA}$, corresponding to noise ranges of $[-300~\mathrm{mV}, 300~\mathrm{mV}]$ and $[-250~\mathrm{mA},250~\mathrm{mA}]$.

\item \textit{ADC quantization:}
Due to the finite ADC resolution, continuous analog signals are mapped to a finite number of discrete digital codes during digitization. This effect is modeled by quantizing the clean simulation data to the nearest ADC code. For a 12-bit ADC with full-scale ranges of $30~\mathrm{V}$ for the output voltage and $25~\mathrm{A}$ for the inductor current, the least significant bit (LSB) values are $30~\mathrm{V}/2^{12}=7.32~\mathrm{mV}$ and $25~\mathrm{A}/2^{12}=6.10~\mathrm{mA}$, respectively. Thus, the quantization error is bounded by $\pm0.5$ LSB, corresponding to $\pm3.66~\mathrm{mV}$ for $v_o$ and $\pm3.05~\mathrm{mA}$ for $i_L$.

\item \textit{Synchronization offset:}
In practical multi-channel sampling systems, $v_o$ and $i_L$ may not be acquired at exactly the same instant, leading to an inter-channel time offset. In this study, the inductor current sampling instant is assumed to lag behind that of the output voltage. The maximum offset is set to $2\%$ of the switching period (i.e., $\tau_{\text{max}} = 0.02T_{\text{sw}}$). This setting gives an upper delay bound of $1~\mu \mathrm{s}$ for the $20~\mathrm{kHz}$ cases. At each sampling instant, an independent random delay $\tau_j\sim\mathcal{U}(0,\tau_{max})$ is applied to the inductor current sample relative to the output voltage sample.

\item \textit{Model--plant mismatch:}
The mathematical model established in Section II captures only the dominant dynamics of the converter, whereas practical converters contain additional unmodeled parasitic effects, such as parasitic inductances and capacitances, arising from device characteristics, packaging, and circuit layout. Although these parasitic effects are often small in magnitude, their omission from the nominal model can introduce model--plant mismatch. The resulting discrepancy may substantially affect the accuracy of parameter estimation. To evaluate this effect, simulation data are generated using a higher fidelity model that incorporates parasitic components as shown in Fig.~\ref{robustness_eval} (b), and the parameters are then estimated using the nominal model.

\end{enumerate}

With different combinations of the above uncertainty sources, the test results under the Case I configuration are summarized in Table~\ref{tab:uncertainty_cases}. Overall, ADC quantization and model--plant mismatch introduce only limited degradation in the estimation accuracy, while measurement noise and synchronization offset have a more noticeable influence due to the direct distortion of the sampled trajectories. The results also show that the parameters in $\boldsymbol{\theta}_1$ (i.e., $C$, $L$, and $R_{\mathrm{load}}$) remain accurately estimated under different uncertainty combinations. In contrast, $R_C$ and $v_F$ are more sensitive to uncertainty, and the individual estimates of $R_L$ and $R_{dson}$ are unreliable due to their strongly coupled effects on the measured states.

Since the imposed uncertainties are stochastic, a Monte Carlo analysis is further conducted to provide a more comprehensive robustness assessment. Specifically, 1000 independent trials are performed under the most severe uncertainty configuration (i.e., ADC-Sync-1\%noise-MPM).
As shown in Fig.~\ref{montecarlo}, the estimation errors of $L$, $C$, and $R_{\mathrm{load}}$ are tightly concentrated, confirming the robustness of the strongly identifiable parameters. The distributions of $R_C$ and $v_F$ are wider than those parameters in $\boldsymbol{\theta}_1$. Nevertheless, even for $v_F$, 95\% of the samples remain within the range of $-11\%$ to $17\%$. For $R_L$ and $R_{dson}$, the individual estimates exhibit large dispersion and cannot be interpreted separately. However, their estimates are strongly correlated along the direction of an approximately constant equivalent resistance, and most samples remain within the $\pm20\%$ band of $R_D=R_L+R_{dson}$. It is worth noting that each training set contains data from only 30 switching cycles, with a sampling frequency of merely twice the switching frequency. Even under this sparse-data condition, the proposed framework remains robust for strongly identifiable parameters and can still provide reliable health indicators for coupled degradation-related parameters under practical uncertainty conditions.

\section{Experiment Verification}

\begin{figure}
  \begin{center}
  \includegraphics[width=3.5in]{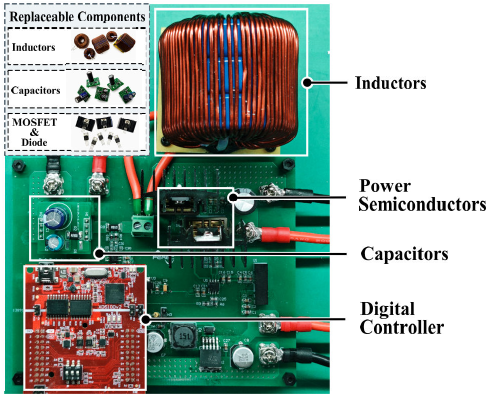}\\
  \caption{Experimental test platform of the buck converter with replaceable components. The inductors, capacitors, and power semiconductors can be readily replaced}\label{buck_board}
  \end{center}
\end{figure}

To validate the proposed method experimentally, a buck converter test platform with replaceable components is developed, as shown in Fig.~\ref{buck_board}. The key components, including the capacitor, inductor, MOSFET, and diode, can be readily replaced to support different hardware configurations. For example, capacitor degradation can be emulated by combining capacitors and series resistors with different values to reproduce aging-induced variations in capacitance and equivalent series resistance (ESR). The required signals are acquired by the onboard sensors using the sampling strategy described in Section III-A. Prior to data collection, the sensing channels are calibrated against oscilloscope measurements to reduce acquisition errors.

\subsection {Experiment Configuration}
Consistent with Section~III, the switching frequency of the buck converter is set to 20 kHz. As illustrated in Fig.~\ref{transient}, transient responses are excited by manually changing the load, and the output voltage and inductor current are recorded during the transient intervals. For each transient response, the recorded window is limited to 30 switching cycles. This setting reflects the short settling interval typically expected in a well-designed converter. The sampling instants are synchronized with the valley and peak of the PWM carrier, corresponding to normalized carrier values of 0 and 1, respectively. Therefore, each transient record contains only 60 sampled points. Similarly, transient data from three load transitions (e.g., $R_0 \to R_1$, $R_1 \to R_2$, and $R_2 \to R_3$) are combined to form the training dataset for one parameter estimation task. The DP simulation framework is configured with the single-step ERK4 scheme and the regularization constraints specified in Section~III-B, with $N_h=16$ for the long-horizon constraint.

In the tests, the degradation behavior of the converter is simulated by replacing component combinations with different values. The baseline values of the passive components (i.e., the inductor and capacitor) are measured using an LCR meter (HIOKI IM3536) at the switching frequency. In particular, to minimize the difference between the actual inductance during operation and the LCR measurement caused by DC bias effects, inductor cores with low permeability are selected. In terms of the active components (i.e., the MOSFET and diode), the values are obtained from static measurements at a specified current. The detailed experimental configurations are summarized in Table~\ref{tab:exp_components}. By systematically interchanging these components, a total of 30 distinct hardware configurations of the converter are formulated for the subsequent experimental validation.

\begin{table}[t]
\caption{Reference Values of Replaceable Components}
\label{tab:exp_components}
\centering
\footnotesize
\setlength{\tabcolsep}{5pt}

\begin{tabular}{c l c}
\hline
Type & Reference value & Test condition\\
\hline
\multirow{5}{*}{\begin{tabular}[c]{@{}c@{}}Capacitor\end{tabular}} 
& $C_1$: $170.42~\mu\mathrm{F}$, $98.05~\mathrm{m}\Omega$
& \multirow{5}{*}{\makecell{at switching frequency\\($20~\mathrm{kHz}$)}} \\
& $C_2$: $141.89~\mu\mathrm{F}$, $59.27~\mathrm{m}\Omega$
& \\
& $C_3$: $158.59~\mu\mathrm{F}$, $101.23~\mathrm{m}\Omega$
& \\
& $C_4$: $144.84~\mu\mathrm{F}$, $156.16~\mathrm{m}\Omega$
& \\
& $C_5$: $94.52~\mu\mathrm{F}$, $256.52~\mathrm{m}\Omega$
& \\
\arrayrulecolor{gray!35}
\specialrule{0.3pt}{0pt}{0.5ex}
\arrayrulecolor{black}
\multirow{2}{*}{\begin{tabular}[c]{@{}c@{}}Inductor\end{tabular}} 
& $L_1$: $333.99~\mu\mathrm{H}$, $268.98~\mathrm{m}\Omega$
& \multirow{2}{*}{\makecell{at switching frequency\\($20~\mathrm{kHz}$)}} \\
& $L_2$: $285.49~\mu\mathrm{H}$, $228.22~\mathrm{m}\Omega$
& \\
\arrayrulecolor{gray!35}
\specialrule{0.3pt}{0pt}{0.5ex}
\arrayrulecolor{black}
\multirow{3}{*}{\begin{tabular}[c]{@{}c@{}}MOSFET\end{tabular}} 
& $M_1$: $160.84~\mathrm{m}\Omega$
& at 3~A ($T_{case}=35.7 ^\circ C $)\\
& $M_2$: $101.25~\mathrm{m}\Omega$
& at 3~A ($T_{case}=35.3 ^\circ C $)\\
& $M_3$: $195.36~\mathrm{m}\Omega$
& at 3~A ($T_{case}=36.8 ^\circ C $)\\
\hline
\end{tabular}
\end{table}

\subsection {Experiment Results Analysis}
As detailed previously, transient data from three load-step tests are combined to perform the parameter estimation task for each hardware configuration. Before the experimental tests, the key replaceable components are characterized off-line to obtain their reference values. However, it is practically inevitable that these offline measurements will exhibit discrepancies compared to the effective parameter values observed during the field operational stage. Particularly, while the passive components are characterized exclusively at the switching frequency offline, their dynamic behavior during actual converter operation is governed by a broader spectral range. Considering DC bias effects and higher-order PCB parasitics, directly evaluating estimation accuracy based on absolute deviations from offline measurements can be misleading.


\begin{table*}[t]
\caption{Percentage Variations of the Estimated Parameters under Different Hardware Configurations}
\label{tab:experiment_analysis}
\centering
\footnotesize
\setlength{\tabcolsep}{3.2pt}
\renewcommand{\arraystretch}{1.05}

\begin{tabular}{@{}c c c c c c c c c @{\quad} c c c c c c c c c@{}}
\toprule
\multirow{2}{*}{Config.}
& \multicolumn{4}{c}{Variation~(\%)}
& \multicolumn{4}{c}{Est. Variation~(\%)}
& \multirow{2}{*}{Config.}
& \multicolumn{4}{c}{Variation~(\%)}
& \multicolumn{4}{c}{Est. Variation~(\%)} \\
\cmidrule(lr){2-5}
\cmidrule(lr){6-9}
\cmidrule(lr){11-14}
\cmidrule(lr){15-18}
& $L$ & $C$ & $R_C$ & $R_D$
& $L$ & $C$ & $R_C$ & $R_D$
&
& $L$ & $C$ & $R_C$ & $R_D$
& $L$ & $C$ & $R_C$ & $R_D$ \\
\midrule

$L_1M_1C_1$ & 0.00 & 0.00 & 0.00 & 0.00 & 0.00 & 0.00 & 0.00 & 0.00
&
$L_2M_1C_1$ & -14.52 & 0.00 & 0.00 & -11.81 & -14.86 & 0.57 & 3.88 & -2.75 \\

$L_1M_1C_2$ & 0.00 & -19.11 & -29.35 & 0.00 & -1.30 & -19.59 & -26.63 & -4.34
&
$L_2M_1C_2$ & -14.52 & -19.11 & -29.35 & -11.81 & -17.41 & -19.75 & -26.53 & -2.68 \\

$L_1M_1C_3$ & 0.00 & -9.78 & 3.24 & 0.00 & 1.57 & -9.77 & 15.47 & -4.12
&
$L_2M_1C_3$ & -14.52 & -9.78 & 3.24 & -11.81 & -14.79 & -8.96 & 10.15 & -4.22 \\

$L_1M_1C_4$ & 0.00 & -17.43 & 59.27 & 0.00 & -1.25 & -19.15 & 49.54 & 2.91
&
$L_2M_1C_4$ & -14.52 & -17.43 & 59.27 & -11.81 & -16.36 & -18.77 & 53.30 & 0.36 \\

$L_1M_1C_5$ & 0.00 & -46.12 & 161.62 & 0.00 & -1.78 & -49.01 & 167.02 & -2.39
&
$L_2M_1C_5$ & -14.52 & -46.12 & 161.62 & -11.81 & -15.84 & -49.17 & 177.52 & -6.16 \\

\addlinespace[0.2em]

$L_1M_2C_1$ & 0.00 & 0.00 & 0.00 & -13.86 & 1.34 & -0.25 & 1.37 & -16.43
&
$L_2M_2C_1$ & -14.52 & 0.00 & 0.00 & -25.67 & -14.07 & 0.26 & -0.79 & -18.60 \\

$L_1M_2C_5$ & 0.00 & -46.12 & 161.62 & -13.86 & -3.45 & -48.92 & 165.06 & -19.57
&
$L_2M_2C_5$ & -14.52 & -46.12 & 161.62 & -25.67 & -18.96 & -48.97 & 160.02 & -22.93 \\

\addlinespace[0.2em]

$L_1M_3C_1$ & 0.00 & 0.00 & 0.00 & 8.03 & 0.92 & 0.27 & 8.27 & 3.91
&
$L_2M_3C_1$ & -14.52 & 0.00 & 0.00 & -3.78 & -14.52 & 0.35 & 8.48 & 2.22 \\

$L_1M_3C_5$ & 0.00 & -46.12 & 161.62 & 8.03 & -2.32 & -48.93 & 180.40 & 4.75
&
$L_2M_3C_5$ & -14.52 & -46.12 & 161.62 & -3.78 & -16.66 & -49.34 & 173.47 & 4.78 \\

\bottomrule
\end{tabular}

\vspace{0.35em}
\footnotesize
\raggedright
\textit{Note:} Only representative configurations are listed. The selected cases include the full capacitor sweep under $M_1$ for both inductors and the boundary capacitor cases $C_1$ and $C_5$ under $M_2$ and $M_3$.
\end{table*}

\begin{figure*}[!t]
  \begin{center}
  \includegraphics[width=\textwidth]{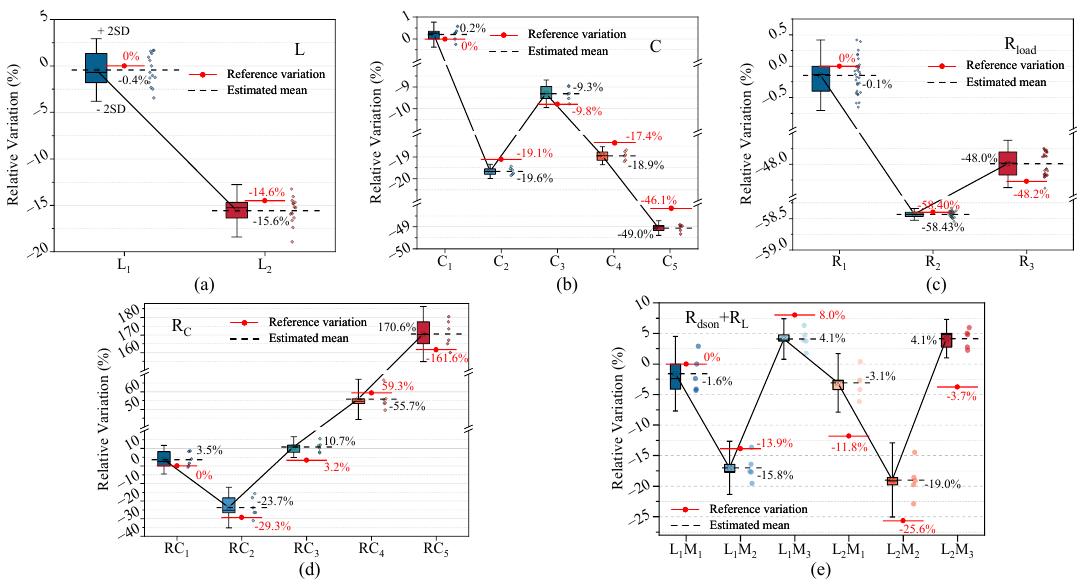}\\
  \caption{Experimental comparison of reference and estimated relative parameter variations under different hardware configurations. (a) Inductance ($L$). (b) Capacitance ($C$). (c) Load resistance ($R_{\mathrm{load}}$). (d) Capacitor ESR ($R_C$). (e) Equivalent resistance ($R_D=R_{dson}+R_L$).}\label{experiment_results}
  \end{center}
\end{figure*}

To mitigate this issue, relative percentage changes are used to characterize health status of the components and evaluate the proposed method. Specifically, the estimated parameters obtained under the configuration $L_1M_1C_1$ are adopted as the reference baseline, fundamentally treating these estimates as the ground truth (i.e., $0\%$ deviation).  It should be noted that the baseline is the average result of three repeated experiments under the same operating conditions for the configuration $L_1M_1C_1$. In this way, the estimated parameters of the other configurations are then normalized to this baseline, and the estimated percentage changes are compared with those derived from the offline measurements. The test results covering multiple component combinations are presented in Table~\ref{tab:experiment_analysis}. Moreover, to provide a more intuitive component-wise comparison, the relative variations of the estimated parameters are further illustrated in Fig.~\ref{experiment_results}.

The comparative experimental results demonstrate that the proposed method can accurately and robustly track the variation trends of the critical components. Several key observations can be readily drawn. First, the relative variations of the inductance $L$, capacitance $C$, and load resistance $R_\mathrm{load}$ can be tracked with high accuracy. Their estimated results exhibit narrow distributions across different hardware configurations, indicating high repeatability, as depicted in Fig.~\ref{experiment_results}(a)-(c). This indicates that the proposed framework successfully captures the dominant dynamic information of the converter and can stably identify the primary parameters. Notably, the estimation distribution for $L$ is slightly wider than that for $C$. This discrepancy can likely be attributed to the practical limitation that current sensors are inherently more susceptible to measurement noise and inaccuracies than voltage sensors.

Second, regarding the capacitor ESR (i.e., $R_C$), the proposed method still captures the degradation trend, as shown in Fig.~\ref{experiment_results}(d). Meanwhile, it is observed that the estimation of $R_C$ exhibits a relatively larger deviation. Specifically, the maximum deviation in relative change reaches 18.78\% under $L_1M_3C_5$, as detailed in Table~\ref{tab:experiment_analysis}. However, this deviation is relatively small compared with the large ESR increase of 161.62\%, especially considering that an ESR value reaching $2\sim3$ times its initial value is commonly used as a capacitor failure criterion. It should be noted that the diode forward voltage $v_F$ is not emphasized in this work. Because its weak contribution to the measured waveforms makes it difficult to identify accurately under measurement noise and model mismatch.

Nevertheless, the estimation of the equivalent resistance $R_D$ is not as accurate as that of the other parameters (e.g., $L, C, R_\mathrm{load}$). The individual estimation results exhibit noticeable dispersion. For instance, as shown in Fig.~\ref{experiment_results}(e), the relative variation rates among different estimation trials for $L_2M_2$ differ by approximately 8\%. Furthermore, there is a considerable discrepancy between the mean estimated values and the reference variations. 

These unsatisfactory results are attributed to a combination of several factors. A primary factor is that the $R_L$ of the inductor and $R_{dson}$ of the MOSFET are not strictly constant. They dynamically fluctuate with temperature and current levels. However, the model derivation for parameter estimation treats the sum of $R_L$ and $R_{dson}$ as a constant, and the reference baseline is acquired from offline static measurements. Another contributing factor is the sampling distortion induced by various sources of uncertainty in the measurement process. It is worth noting that the estimated variation of $R_D$ generally follows the reference trend. This suggests that utilizing $R_D$ as a condition indicator for degradation monitoring remains practically meaningful, particularly for detecting severe fault scenarios such as a multifold increase in the MOSFET on-state resistance caused by gate drive anomalies. 

\subsection {Discussion}
The proposed method reliably tracks the critical parameters of the dc--dc buck converter under practical hardware conditions, and provides meaningful condition indicators. As a representative case study, the concept of DP shows considerable potential and application value in the field of power electronics.
\begin{enumerate}
\item The framework establishes a differentiable mapping between critical component parameters and externally measured converter waveforms. This implies that the proposed framework is not confined to condition monitoring, but may also be extended to other applications, such as converter design optimization and model predictive control.

\item It introduces a novel perspective by treating the numerical solver as a differentiable computational graph to simulate converter dynamics. This formulation naturally offers the potential for integration with neural-network-based deep learning techniques.

\item It provides a topology-general framework based on numerical time-stepping simulation, which can be extended to other converter topologies, such as dc--ac inverters and resonant converters.

\end{enumerate}

Although the DP simulation framework yields promising results, several fundamental challenges remain to be addressed from the perspective of condition monitoring to facilitate its widespread industrial implementation.

\begin{enumerate}
\item Component characteristics are highly sensitive to operating conditions such as temperature and load current. 
Condition monitoring often reduces such behavior to a fixed estimated value and compares it with an offline static reference. A more appropriate health-state representation is needed to account for such a mismatch.

\item Certain parameters exhibit limited identifiability because of their weak sensitivity to measured states and strong mutual coupling. A rigorous mathematical framework is needed to systematically analyze and quantify parameter identifiability, rather than relying solely on empirical observations from experiments.

\item Uncertainty sources, such as measurement noise and synchronization error, can affect the credibility of parameter estimation. These uncertainties should be explicitly modeled and propagated through the estimation process, rather than being considered only as perturbations for robustness testing.

\end{enumerate}

\section{Conclusion}

This article has presented a DP simulation-based parameter estimation method for power electronic converters. By reformulating numerical time-stepping simulation as a differentiable computational graph, the proposed DP simulation framework enables end-to-end gradient-based optimization of converter parameters directly through the physical evolution dynamics. Moreover, it seamlessly maintains gradient propagation across converter topological transitions and ensures reliable model training even with sparsely sampled data. The effectiveness of the proposed method has been verified through both simulation and experimental studies on a dc–-dc buck converter. The dominant parameters, including $L$, $C$, and $R_{\mathrm{load}}$, are accurately estimated, with the maximum experimental variation deviation kept within 5\%, while the capacitor ESR $R_C$ and the equivalent resistance $R_D$ show maximum deviations within 20\% and can still serve as useful health indicators for maintenance guidance. Owing to its flexible formulation, the framework can be readily combined with well-established numerical integration schemes and exhibits immense potential for integration with rapidly advancing deep learning techniques. It is expected that this physically interpretable framework will facilitate the broader application of physics-informed machine learning in power electronics.


%





\ifCLASSOPTIONcaptionsoff
  \newpage
\fi





\bibliographystyle{IEEEtran}
\bibliography{IEEEabrv,Bibliography}

\vfill


\end{document}